\documentclass[a4paper,10pt]{article}

\usepackage{color}
\usepackage{graphicx}
\usepackage{amsmath}
\usepackage{graphics}
\usepackage{amsthm}
\usepackage{amsfonts}
\usepackage{url}
\usepackage{multirow}
\usepackage{rotating}
\usepackage{wrapfig}
\usepackage{authblk}
\usepackage[bookmarks=false]{hyperref}

\newtheorem{definition}{Definition}

\newcommand{\LV}{Low Voltage }
\newcommand{\MV}{Medium Voltage }
\newcommand{\HV}{High Voltage }
\newcommand{\CPL}{Characteristic Path Length }
\newcommand{\APL}{Average Path Length }
\newcommand{\bet}{betweenness }
\newcommand{\PG}{Power Grid }
\newcommand{\etal}{\textit{et al.\ }}
\newcommand{\G}{Grid }
\newcommand{\sw}{Small-world }
\newcommand{\MLV}{Medium and Low Voltage }
\newcommand{\CNA}{Complex Network Analysis }

\newcommand{\Keywords}[1]{\par\noindent
{\small{\em Keywords\/}: #1}}

\begin{document}
\title{Towards Decentralized Trading: A Topological Investigation of the Dutch Medium and Low Voltage Grids}

\author{Giuliano~Andrea~Pagani} 
\author{Marco~Aiello}

\affil{Distributed Systems Group\\Johann Bernoulli Institute for Mathematics and Computer Science
\\University of Groningen\\ Groningen, The Netherlands\\
\vspace{0.3cm}
email: \url{{g.a.pagani,m.aiello}@rug.nl}\\
 \url{http://www.cs.rug.nl/ds/}
}

\maketitle

\begin{abstract}
The traditional \PG has been designed in a hierarchical
fashion, with energy pushed from the large scale production factories
towards the end users. With the increasing availability of micro
and medium scale generating facilities, the situation is
changing. Many end users can now produce energy and share it over the
Power Grid. Of course, end users need incentives to do so and
 want to act in an open decentralized energy
market. In the present work, we offer a novel analysis of the Medium
and Low Voltage Power Grids of the North Netherlands using statistical
tools from the Complex Network Analysis field. We use a weighted model
based on actual Grid data and propose a set of statistical measures to
evaluate the adequacy of the current infrastructure for a
decentralized energy market. Further, we use the insight gained by the
analysis to propose parameters that tie the statistical topological
measures to economic factors that influence the attractiveness
of participating in such decentralized energy
market, thus identifying the important topological parameters
to work on to facilitate such open decentralized markets.
\end{abstract}


\section{Introduction}
The \PG is one of the engineering masterpieces
 of the XIX-XX century, being one of the most important infrastructures
that contributes to the economic growth and welfare of any country. It
has been designed as a hierarchical system with large
generating facilities on top, and a pervasive network of cables to
distribute the energy to the geographically
dislocated end users. Traditionally, 
it has been created to be managed by a
monopolist or an oligarchy of actors. Typically, energy availability is given for granted, though
its importance becomes well too apparent both at the household and
country level when prolonged blackouts strike and electricity flow is
interrupted~\cite{Anderson2007,Balducci2002}. 

Something is though changing in the way energy is produced and
distributed due to both technological advancements and the
introduction of new policies. A clear trend of market unbundling is
emerging (e.g.,~\cite{cossent09,joskow08}) entailing
the addition of many new players to the
energy sector with the possibility to produce, sell and distribute
energy. From the technological perspective, new energy generation
facilities (mainly based on renewable
sources) are becoming widely accessible. These are convenient and available at both the
industrial and the residential
scale~\cite{mar:mic06,lovins02}. The term \emph{Smart Grid}, which does not yet have a unique
agreed definition, is sometimes used to define the new scenario of a
Grid with a high degree of delocalization in the production and trading of
energy. The new actors, who are both producers and consumers of
energy, referred to as  {\em prosumers}, are becoming more numerous and
will most likely demand a market with total freedom to energy trading~\cite{vai:pow05}. 
In this coming scenario the main role of the \HV \G may change, while
the Distribution \G (i.e., \MV and \LV end of the Power Grid) gain
more and more importance, while requiring a major update. In fact, the energy interactions between
prosumers will increase and most likely occur at a rather local level,
therefore involving the Low and Medium Voltage Grids. 

Given such emerging scenario, we propose to look at the lower layers
of the \PG in a statistical manner, considering global metrics
coming from the field of Complex Network Analysis. Few such studies
exist in the literature using unweighted models of the \HV
Grids. These have been performed especially to establish the
resilience to failures of critical national infrastructures. It will
not surprise that often such studies have appeared immediately after a
major national blackout. More recently, these techniques have also
been used for generating tests for the Smart Grid in~\cite{Wang2010}.
Here we propose a novel study of the
properties of the Medium and \LV networks using the Northern part of the
Netherlands as data source.  Our investigation goes beyond the study
of the existing as it also proposes a way of evaluating the
infrastructure in terms of its ability to support delocalized energy
trading. We argue that global topological statistical measures do
influence the eagerness of prosumers to trade energy, which in turn,
 entails a structural modification of the Power Grid.

The paper is organized as follows. Motivation for the present study and an initial overview
of the state of the art are presented in Section~\ref{sec:SoArt}. In
Section~\ref{sec:analysis}, we introduce the data set used for the study and
provide essential definitions. In Section~\ref{sec:unweighted} we provide an initial unweighted Complex Network Analysis. Section~\ref{sec:weighted} is dedicated to a weighted Complex Network Analysis which
provides the best insight on the available sample and a set of novel
measures in the statistical study of the Power Grid. A comparison with an unweighted \CNA is provided
Section~\ref{sec:unweightedweightedcomparison}. The proposal on how to
tie topological metrics to economic factors is presented in
Section~\ref{sec:economic}, together with an example. A detailed
account of related work on Complex Network Analysis for the (High
Voltage) \PG is provided in Section~\ref{sec:relwks}. 
Final discussion and conclusions are offered in Section~\ref{sec:conc}.

\section{A Statistical View on the Power Grid}\label{sec:SoArt}

Many scientific knowledge areas contribute to the design
and analysis of Power systems: Physics (electromagnetism, classical
mechanics), Electrical engineering (AC circuits and phasors, 3-phase
networks, electrical systems control theory) and Mathematics (linear
algebra, differential equations). Such traditional studies 
tend to have a ``local'' view of the Grid, e.g., defining how to
design a transformer and predicting its functioning. Typically studies
tend to focus on the physical and electrical properties
(e.g.,~\cite{Anghel06}), or the characteristics of the \PG as a
complex dynamical system~\cite{Dobson012}, or again, the
control theory aspects~\cite{Hiskens97}.

The move from a ``local'' to a ``global'' view of the \PG as a complex
system is possible by resorting to statistical graph theory, better known as
{\em Complex Network Analysis (CNA)}, that is, a statistical analysis of the
dynamics of large graphs with the goal of identifying their characterizing
properties, such as the average path length between any two
nodes. Studies of this sort have been recently performed for example
on the American ~\cite{Chassin05,Albert04}, the
Italian~\cite{Crucitti04}, and the Scandinavian~\cite{holmgren06} Grids; the European Grid as a whole is analysed
in~\cite{casals07,Rosas-Casals2009}. Interestingly, these studies,
though using the same mathematical machinery, reach different
conclusions on the characteristics of the identified graph (e.g., node degree
distribution, network category). An explanation for this is that
different geographical infrastructures may indeed have different
layouts, thus yielding different topological properties.
%

Motivations for such CNAs are common to all previous researches and are mainly two:
identifying the right complex network model for the Power Grid;
and provide a resilience analysis for predicting blackouts
and critical components of the infrastructure. 
Folkloristic is that \PG CNA studies are 
`popular' after a major blackout occurs, such as the  North
  American black-out of
  2003\footnote{\url{http://news.bbc.co.uk/2/hi/americas/3152451.stm}}
  or the Italian one of
  2003\footnote{\url{http://news.bbc.co.uk/2/hi/3146136.stm}} (e.g.,~\cite{Albert04,Crucitti2004a,Crucitti04,Chassin05}). 
The fragility of the \PG has been the major reason of concern that has
determined the focus of such CNA studies on the
High Voltage.
In fact, High Voltage failures impact a large part of the network, thus resulting in electricity service disruptions of large portions of a whole country. From a graph perspective, the \HV samples for an entire country are usually quite small in terms of order and size.
For instance, the Grids connecting the 15 European countries analysed by
Rosas-Casals \etal in~\cite{Rosas-Casals2009} are composed overall of about
2700 nodes and more than half of the analysed samples are below 100
nodes; moreover, the graphs studied in~\cite{Rosato2007,Crucitti2005}
are well below 200 both nodes and edges, while the sample analysed
in~\cite{Crucitti04} has less than 400 nodes and a little more
than 500 edges.

The motivation for the current work is quite different from previous approaches. We consider
the \PG as an infrastructure for decentralized energy exchange. To this end, we are interested in the properties of
the Medium and Low Voltages. Furthermore, we consider the
\PG not only in its basic topology to assess resilience and
connectivity, but also in the physical properties of the lines to assess
the capacity of the infrastructure in supporting distribution. To the
best of our knowledge, the study has novel motivations, in line
with the current trends of the Smart Grid, but also studies new
networks with novel weighted models.

\section{Northern Netherlands Medium and Low Voltage Complex Network Analysis}
\label{sec:analysis}

\begin{figure}[htbp]
\centering
\resizebox{6cm}{!}{ \includegraphics{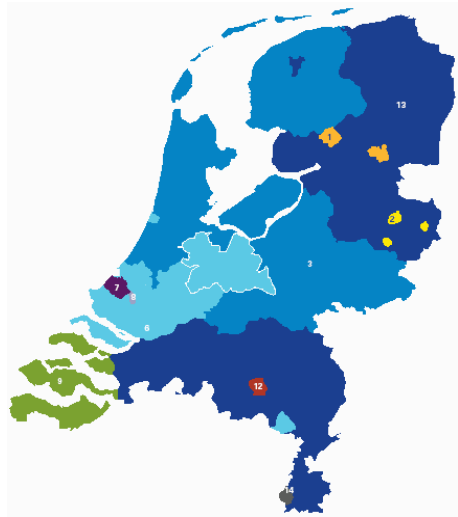}}
\caption{Distribution companies over the Netherlands. Each color corresponds to one company. {\emph{(Source: www.energieleveranciers.nl)}}}
\label{fig:distrib}
\end{figure}

We focus on the Medium and Low Voltage \PG networks of (Northern) Netherlands. This choice is dictated by the fact that the Netherlands has a modern infrastructure.
The Dutch \HV \PG is owned and managed by one player, Tennet, while the lower layers are partitioned geographically among fourteen companies\footnote{Zone identifier and distribution provider: 1) RENDO Netwerken, 2) Cogas Infra en Beheer, 3) Liander (former Continuon Netbeheer), 6) Stedin (former Eneco), 7) Westland Infra, 8) ONS Netbeheer (now Stedin), 9) DELTA Netwerkbedrijf, 12) NRE Netwerk, 13) Enexis (former Essent Netwerk), 14) InfraMosane (now Enexis).} that have their own distribution network across the country. The partition of the territory among energy distribution companies is shown in Figure~\ref{fig:distrib}.

\begin{table*}[htb]
\begin{center}
\begin{small}
    \begin{tabular}{| p{0.4cm} || p{0.7cm} | p{0.5cm}  | p{0.9cm} | p{0.9cm}  | p{0.9cm} |p{1.0cm} || p{1.0cm}| p{1.0cm} |p{1.0cm} |}
    \hline
& \multicolumn{6}{|c||}{\sc{Present study}} &  \multicolumn{3}{|c|}{\sc{Random Graph}}\\
\hline
    ID & Order & Size & Avg. $d$ & APL & CPL & $\gamma$ &  APL &  CPL &  $\gamma$\\ \hline\hline
    1 & 17 & 18 & 2.118 & 3.398 & 3.313 & 0.00000 & 1.427 & 1.688 & 0.13726\\ \hline
    2 & 15 & 16 & 2.133 & 3.086	& 3.000&0.00000 & 2.319 & 2.358 & 0.00000\\ \hline
    3 & 24 & 23 & 2.087 & 4.499 & 4.228&0.00000 & 3.127 & 3.091 & 0.05508\\ \hline
    4 & 30 & 29 & 1.933 & 4.545 & 4.449&0.00000& 1.860 & 2.242 & 0.05778\\ \hline
    5 & 188 & 191 & 2.032 & 17.726 & 17.878&0.00000& 3.846 & 4.345 & 0.00532\\ \hline
    6 & 10 & 9 & 1.800 & 2.423 & 2.223&0.00000& 0.978 & 1.167 & 0.26667\\ \hline
    7 & 63 & 62 & 1.968 & 5.204 & 5.404&0.00000& 2.514&	2.904&	0.03175\\ \hline	
    8 & 28 & 27 & 1.929 &	4.784 &	5.000&0.00000& 2.553&	2.945&	0.04762\\ \hline
    9 & 133 & 140 & 2.105 &	11.543&	11.366&0.01112& 3.702 & 4.172 & 0.01482\\ \hline
    10 & 124 & 138 & 2.226&	8.053&	7.070&0.00869& 3.010 & 3.540 & 0.02914\\ \hline
    11 & 31 & 30 & 1.935&	4.353&	4.357&0.00000& 1.590 & 1.969 & 0.07475\\ \hline
    \end{tabular}
    \end{small}
\end{center}

\caption{Low Voltage samples from the northern Netherlands \PG compared with Random graphs of the same size.\label{fig:uLow}}
\end{table*}

The Grid information used in this study is provided courtesy of Enexis
B.V., the distribution operator of Northern Netherlands. The provided
data includes information about the transformers in the Grid together
with the distribution substations. The data set also provides
information about the distribution lines used to connect substations
containing the length of cable and other interesting physical properties (e.g., resistance, capacity, voltage). For the sake of
precision, we define the notion of a (Weighted) Power Grid graph.
\begin{itemize}
 \item All the substations and transformers are considered equal and are represented as nodes of the graph.
\item The cables connecting the substations are considered equal despite the differences in voltages and current carried and their physical properties, and thus modelled as unweighted edges in the graph.
\item For the data samples that present disconnected components, each component is treated as a distinct graph. 
\item The edges are considered undirected.
\end{itemize}
These assumptions are common in Power Grid analysis from a graph theoretic perspective, see for instance~\cite{Crucitti2004a,casals07,Watts98,Rosato2007,Crucitti04,holmgren06} and lead to the following definition.

\begin{definition}[Power Grid graph]
A {\em Power Grid graph} is a graph $G(V,E)$ such that each element 
$v_i \in V$ is either a substation, transformer, or consuming unit of a
physical Power Grid. There is an edge $e_{i,j}=(v_i,v_j) \in E$
between two nodes if there is physical cable connecting directly the elements represented by $v_i$ and $v_j$.
\end{definition}

The next step is to bring cable properties into the graph definition.
\begin{itemize}
 \item For each cable connecting elements in the \G a weight is defined based on the multiplication of the following quantities:
\begin{itemize}
 \item The principal resistance characterizing the cable (whose value is given in Ohm/km).
\item The length of the cable (whose value is given in km).
\end{itemize}
\item A special kind of connection is defined in the \PG known as a
  `link'. These are connections, usually very short, with negligible
  resistance for which the specific value is not provided in the
  dataset. For edges representing these links a conventional weight of $10^{-9}$ is given. This does not affect the overall validity of the weighted model since the number of links in a sample is extremely limited (about 1\% of the overall connections are made of links).
\end{itemize}

\begin{definition}[Weighted Power Grid graph]\label{def:wpgg}
A {\em Weighted Power Grid graph} is a Power Grid graph $G_w(V,E)$ with an additional function $f:E \to \mathbb{R}$ associating a real number to an edge representing the resistance, expressed in Ohm, of the physical cable represented by the edge.
\end{definition}

The analysis we perform uses samples from the \LV and \MV Grids. The
\LV samples sum up to a total of 663 nodes and a 683 edges; while the
\MV samples sum up to 4185 nodes and a 4574 edges. The size of the
data set, tough being a sample and not the whole network, is about the same size or larger than those used in other available studies on the
(High Voltage) Power Grid
~\cite{Watts98,Strogatz2001,Crucitti04,holmgren06,Rosas-Casals2009,casals07,Rosato2007}. We
begin our analysis by considering the unweighted model to
derive basic topological properties and then proceed with a richer
investigation by introducing graph weights.

\begin{table*}[htb]
\begin{center}
\begin{small}
    \begin{tabular}{| p{0.4cm} || p{0.7cm} | p{0.5cm}  | p{0.9cm} | p{0.9cm}  | p{0.9cm} |p{1.0cm} || p{1.0cm}| p{1.0cm} |p{1.0cm} |}
    \hline
& \multicolumn{6}{|c||}{\sc{Present study}} & \multicolumn{3}{|c|}{\sc{Random Graph}}\\
\hline
    ID & Order & Size & Avg. $d$ & APL & CPL & $\gamma$ &  APL &  CPL &  $\gamma$\\ \hline\hline
 1 & 191 & 207 & 2.168&	9.288&	8.990&0.00296& 4.616 & 5.079 & 0.00225  \\ \hline
2 & 884 & 1059 & 2.396 & 9.817 & 9.527&0.00494& 5.440 & 6.010 & 0.00170\\ \hline
3 & 444	& 486 & 2.189 &	11.033	&10.858&0.00537& 5.547 & 6.163 & 0.00333\\ \hline
4 & 472	& 506 & 2.144 & 17.095 & 17.174&0.01360& 5.039 & 5.700 & 0.00106\\ \hline
5 & 238	&245		&2.059	&11.715	&11.580&0.00000& 3.558 & 4.234 & 0.00595\\ \hline
6 & 263	&288&		2.190	&12.775	&12.311&0.01118& 5.046 & 5.368 & 0.01080\\ \hline
7 & 217	&229&		2.111	&10.321	&10.241&0.00140& 4.894 & 5.391 & 0.00121\\ \hline
8 & 366	&382&		2.087	&15.113	&14.546&0.00000 & 4.691 & 5.249 & 0.00405\\ \hline
9 & 218	&232&		2.128	&10.850	&10.915&0.00000& 5.454 & 5.856 & 0.00539\\ \hline
10 & 201&	204		&2.030	&15.742	&15.257&0.00166& 4.898 & 5.503 & 0.00491\\ \hline
11 & 202&	213		&2.109	&13.504&	12.891&0.00140&4.801&	5.217&	0.08750\\ \hline
12 & 25	&24	&	1.920	&5.781	&5.500&0.00000& 4.924&	5.084&	0.00000\\ \hline
13 & 464	&499		&2.151	&13.144	&12.703&0.00036& 4.718 & 5.390 & 0.00209\\ \hline
    \end{tabular}
    \end{small}
\end{center}

\caption{Medium Voltage samples from the northern Netherlands \PG compared with Random graphs of the same size.\label{fig:uMed}}
\end{table*}

\section{Unweighted \PG study}\label{sec:unweighted}

The typical study of the \PG as a complex system considers \HV
samples for identifying how fragile the infrastructure is. We use
similar techniques for the Medium and Low voltage. Let us begin by
recalling the basic complex network quantities.
\begin{definition}[Adjacency, neighbourhood and degree]
 If $e_{x,y} \in E$ is an edge in graph $G$, then $x$ and $y$ are {\em adjacent,} or {\em neighbouring,} vertices, and the vertices $x$ and $y$ are \emph{incident} with the edge $e_{x,y}$.
The set of vertices adjacent to a vertex $x \in V$, called the
\emph{neighbourhood} of $x$, is denoted by $\Gamma(x)$. The number
$d(x)=|\Gamma(x)|$ is the \emph{degree} of $x$.
\end{definition}
A global measure for a graph is given by its average distance among
any two nodes.
\begin{definition}[Average path length (APL)]
Let $v_i \in V$ be a vertex in graph $G$, the \emph{average path length} for $G$, $L_{av}$ is:
\[
 L_{av} = \frac{1}{N \cdot (N-1)} \sum_{i\neq j}d(v_i,v_j)
\]
where $d(v_i,v_j)$ is the finite distance between $v_i$ and $v_j$ and $N$ is the order of $G$.
\end{definition}
\begin{definition}[Characteristic path length (CPL)]
Let $v_i \in V$ be a vertex in graph $G$, the \emph{characteristic path length} for $G$, $L_{cp}$ is defined as the median of ${d_{v_i}}$ where:
\[
 d_{v_i} = \frac{1}{(N-1)} \sum_{i\neq j}d(v_i,v_j)
\]
is the mean of the distances connecting $v_i$ to any other vertex $v_j$ in $G$ and $N$ is the order of the $G$.\\
\end{definition}
A measure of the average `density' of the graph is given by the
clustering coefficient, characterizing the extent to which vertices
adjacent to any vertex $v$ are adjacent to each other.
\begin{definition}[Clustering coefficient (CC)]
The {\em clustering coefficient} $\gamma_v$ of $\Gamma_v$ is
\[
\gamma_v=\frac{|E(\Gamma_v)|}{\binom{k_v}{2}} 
\]
where $|E(\Gamma_v)|$ is the number of edges in the neighbourhood of
$v$ and $\binom{k_v}{2}$ is the total number of $possible$ edges in
$\Gamma_v$.
\end{definition}
This local property of a node can be extended to an entire graph by
averaging over all nodes of the graph.

\subsection{Basic analysis}

We now consider these classic measures on the data of the Dutch
Power Grid. We divide our data
set in samples of topologically connected regions. In
Table~\ref{fig:uLow}, we report the basic analysis on the data
modelled as an unweighted graphs and we compare each sample belonging
to \LV network with a random
graph of the same size  and order. The analysis for
the \MV is reported in Table~\ref{fig:uMed}. Referring to
the table, the first column is the ID of the sample, the second and
third represent the number of vertices $N$ (order) and edges $M$
(size), respectively. The average degree (fourth column) is defined as
$<\!k\!>=\frac{2M}{N}$. The fifth and sixth columns report the average and
characteristic path lengths, that is the average of the minimum
distance between any two given nodes and the median of the same
quantities, respectively. The seventh column provides an indication of
the clustering coefficient of the nodes, that is, broadly speaking,  an average value of
the power of a node to participate in connected aggregation with other
nodes close to it. 

We remark that the average node degree does not have highly different
values in the Low and Medium Voltage samples, they are both around
2. Computing the mean over all samples' average node degree gives a
value of $\overline{<\!k\!>}=2.074$ with a very small variance
$\sigma_{<k>}=0.017$. This value appears to be almost constant
considering the \LV and \MV samples since the variance of the two
categories is even smaller ($\sigma_{<k>_{LV}}=0.016$,
$\sigma_{<k>_{MV}}=0.012$). An almost constant average degree is also
characteristic of the \HV Power Grid~\cite{casals07}, though with a
slightly higher value $<\!k\!>\cong2.8$. This limited number of edges a
node can manage can be regarded as a physical limit that each \PG
substation has to satisfy.

Considering path measures: \APL  and \CPL of the \LV segment of
the network have generally a smaller path length compared to the \MV one. The clustering coefficient is very small especially for
the \LV network for which many samples have a zero value (i.e.,
absence of triangles in the graph). The difference in path
length between the \LV and \MV is due to the higher number of nodes
the \MV network samples have while holding the same average
node degree as the Low Voltage, together with the absence of long
distance edges. This implies a longer path to connect any two nodes in
a bigger network. In addition, these values of APL and CPL are in
general quite high, if compared to other networks such as the World
Wide Web.

The clustering coefficients for the Low Voltage segment of the network
are generally small; this is due to the strong hierarchical
design of this layer of the physical network which resemble a
tree-like structure. Contrarily, the Medium Voltage segment generally
presents higher values for the clustering coefficient, this can be
justified by the different purpose the Medium Voltage network has in
which meshed components and connection redundancies are much more
likely to be present for robustness reasons.

To gain an even better understanding of the tables just presented it
is useful to compare the numbers obtained with those of Random graphs~\cite{Erdos1959} and to
identify the possible presence of {\em Small-world}
properties. Small-world networks (SW), proposed by Watts and Strogatz in~\cite{Watts98},
own two important aspects at the same time: characteristic path length
close in value to the one of a random graph (RG) ($CPL_{SW} \approx
CPL_{RG}$) but a much higher clustering coefficient ($CC_{SW} \gg
CC_{RG}$). Small-worlds are a better model than random graphs for social networks and
other phenomena and thus a candidate for modelling the \PG too.
To make the comparison genuine, random graphs are generated with the
same number of nodes and edges as the real samples, imposing the
resulting graphs not to have disconnected components. The values are
presented on columns eight to ten of Tables~\ref{fig:uLow}
and~\ref{fig:uMed}. We note how the CPL of the Grid samples is on
average twice as big as the random generated samples, thus comparable
to  as the definition of Small-world graph according
to~\cite{Watts98}. In addition the clustering coefficient of the Grid samples is almost
always smaller than the result obtained for the random
generated samples; this completely contradicts the definition of
Small-world graph according to~\cite{Watts98}. Watts and Strogatz
~\cite{Watts98}  impose the following condition to the graphs they
study: $N \gg k \gg ln(N) \gg 1$ where $N$ is the number of nodes, $k$
is the number of edges per node. Such a condition is not satisfied by
the Northern Netherlands samples and generally it is not satisfied by
\PG networks as pointed by Wang \etal in
\cite{Wang2010}. Interestingly, the same condition is also not
satisfied by the Western States \HV Power Grid Watts and Strogatz use
in ~\cite{Watts98} and Watts analyses in ~\cite{Watts03}, while the
results for CC and CPL satisfy the conditions for a Small-world
networks. Another study (i.e., ~\cite{casals07}) considering the
European \HV \PG shows that the \sw phenomenon is not shown by all the
considered Grids, since especially the smaller (in terms of order and
size) Grids fail to satisfy CC condition.

In summary, the Northern Netherlands \MV and \LV samples show a very
small value of average node degree. This is mainly
independent from the size and the different purpose of the network being almost constant despite
the different samples considered. In addition, the path
length is quite high, given the order of the graphs, compared with
other types of complex networks e.g., the World Wide Web. This
relative high path length together with very small clustering
properties suggests that the networks analysed do not strictly follow
the definition of Small-world or, in terms of decentralized energy
negotiation, it suggests that perhaps a structural change to decrease
path length (especially the weighted one) might be necessary to empower
delocalization. We provide an initial proposal in Section~\ref{sec:economic} on
how to achieve this.

\subsection{Node Degree Distribution}

To have a general understanding of the overall characteristics of a
network it is useful to compute certain statistical measures, one of
which is the node degree probability distribution.
More formally,
\begin{definition}[Node degree distribution]\label{def:ndd}
Consider the degree $k$ of a node in a graph as a random variable, the
function 
\[
N_k=\{v\in G:\: d(v)=k\}
\]
is called {\em node degree distribution}.
\end{definition}
The shape of the distribution is a salient characteristic of the
network. For the Power Grid,
the shape is typically either exponential or a power-law.
More precisely an exponential node degree ($k$) distribution has a
fast decay in the probability of having nodes with relative high node
degree and follows a relation:
\[
 P(k)=\alpha e^{\beta k}
\]
where $\alpha$ and $\beta$ are parameters of the specific network
considered. While a power-law distribution has a slower decay with higher probability of having nodes with high node degree:
\[
 P(k)=\alpha k^{-\gamma}
\]
where $\alpha$ and $\gamma$ are parameters of the specific network considered.

Power-law distributions are very common in many real life networks
both created by natural processes (e.g., food-webs, protein
interactions) and by artificial ones (e.g., airline travel routes,
Internet routing, telephone call graphs),~\cite{bar:lin03}. Having a power-law distribution for node
degree means that few nodes have a very high degree and the majority
of nodes have very small degree. The types of networks that follow this property are referred as Scale-free networks (\cite{Barabasi2000,Barabasi2009,Albert2000}); typical
examples of Scale-free networks are the World Wide
Web, the Internet, metabolic networks, airline routes and
many others. From the dynamic point of view, these networks are
modelled by a preferential attachment model, that is, when new nodes
and arcs are added to a graph, these are more likely to connect to
nodes which have already a high degree,~\cite{Barabasi1999,AielloW2002}.
In addition this network structure provides special
reliability properties: high degree of tolerance to random failures
and, at the same time, very sensitive to targeted attacks towards 
hubs~\cite{Albert2000,Moreno03,Crucitti2004a}.

We compute the node degree distribution for every sample both for \LV
and \MV segments. For the most significant samples i.e., those
belonging to the \MV and the big ones belonging to the \LV part the
node degree cumulative distribution seems to follow a power-law: $P_k
\sim k^{-\gamma}$. A prompt method to investigate if
the node degree follows a power-law is to plot the cumulative node
degree distribution on a log-log scale~\cite{Amaral2000}. 
If the distribution in a log-log plot follows a
straight line the distribution can be considered a power-law, while if
the decay is faster this might indicate an exponential
distribution. It is also possible to apply data fitting techniques
(e.g., non-linear least square method) to identify the $\gamma$
parameter of a power-law. 

\begin{figure}
   \centering
   \includegraphics[width=\textwidth]{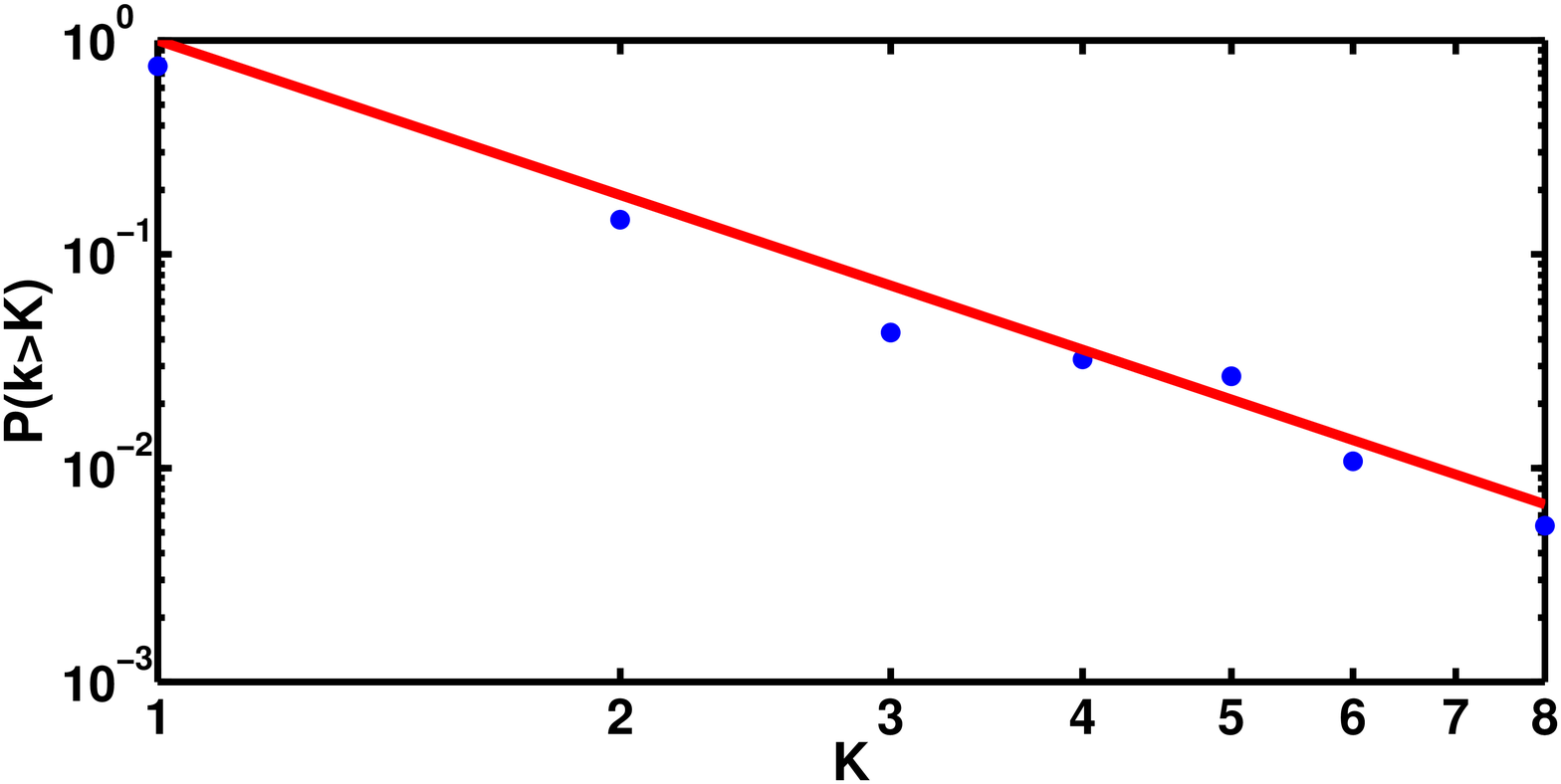}
   \caption{Node Degree Cumulative Probability Distribution for Low Voltage sample \#5. Circles represent sample data, while straight line represents a power-law with $\gamma=2.402$.}
\label{fig:nddLV5}

  \centering
   \includegraphics[width=\textwidth]{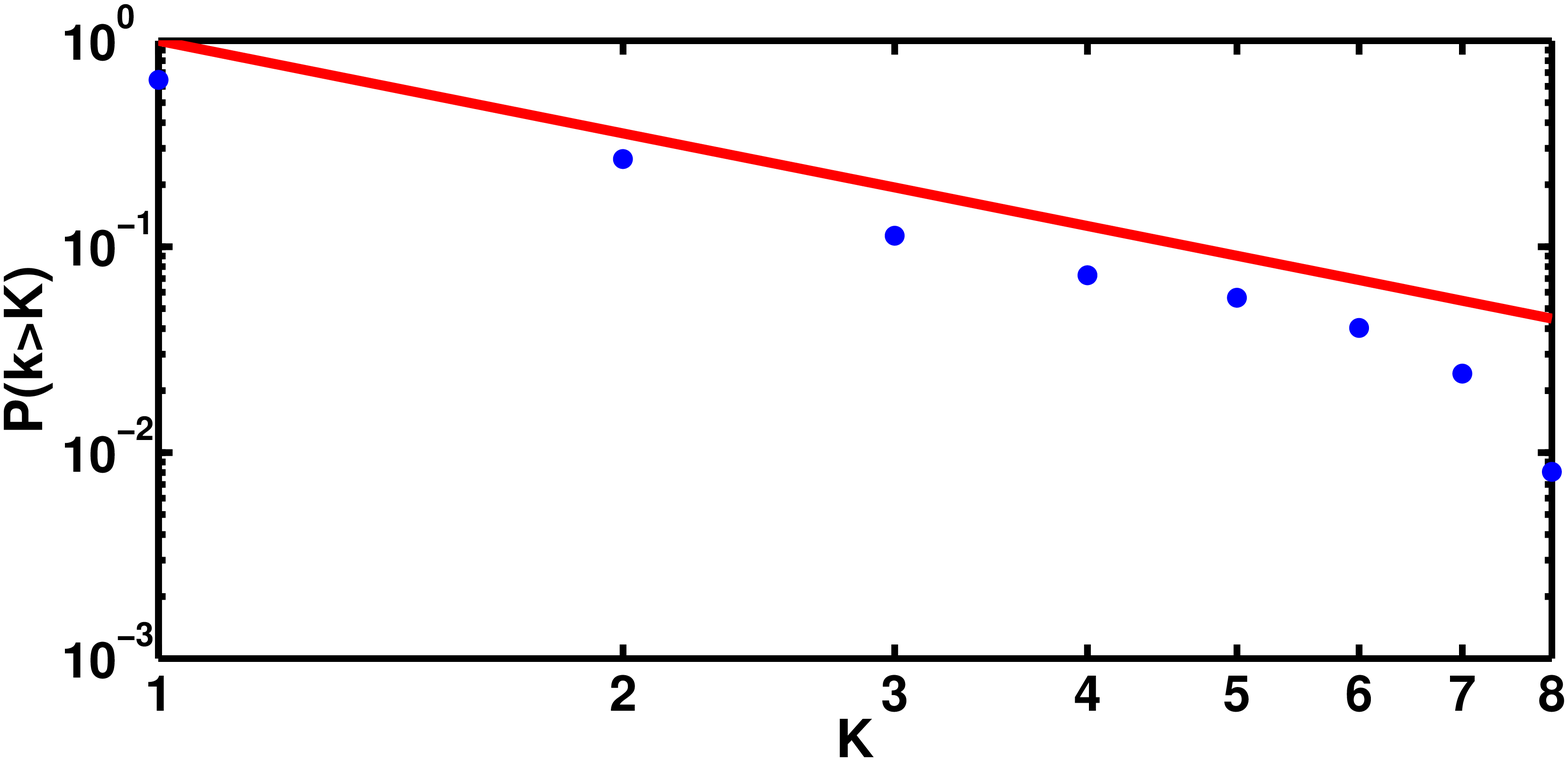}
   \caption{Node Degree Cumulative Probability Distribution for Low Voltage sample \#10. Circles represent sample data, while straight line represents a power-law with $\gamma=1.494$.}
\label{fig:nddLV8}
\end{figure}

\begin{figure}
   \centering
   \includegraphics[width=\textwidth]{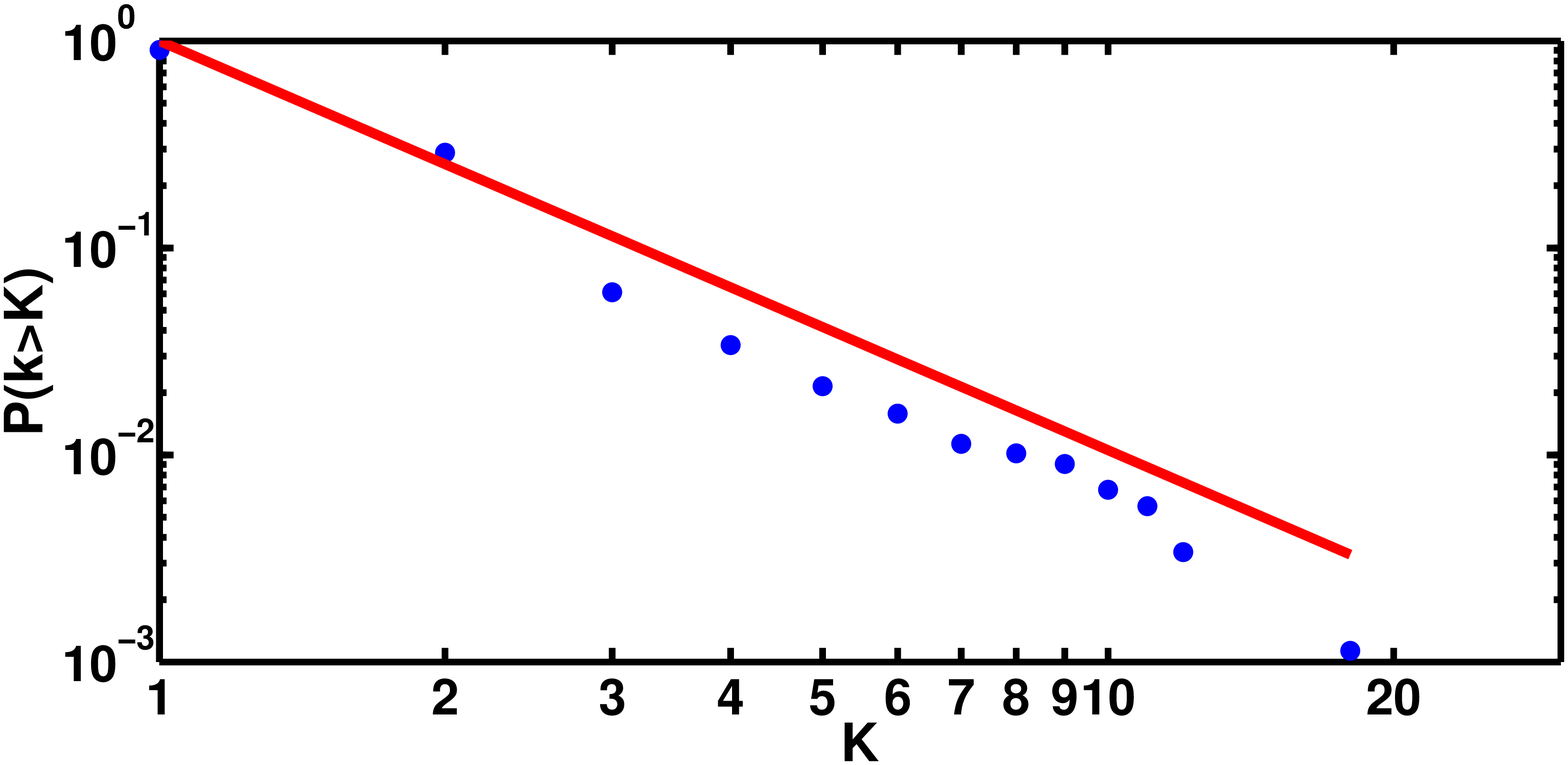}
   \caption{Node Degree Cumulative Probability Distribution for Medium Voltage sample \#2. Circles represent sample data, while straight line represents a power-law with $\gamma=1.977$.}
\label{fig:nddMV2}
  \centering
   \includegraphics[width=\textwidth]{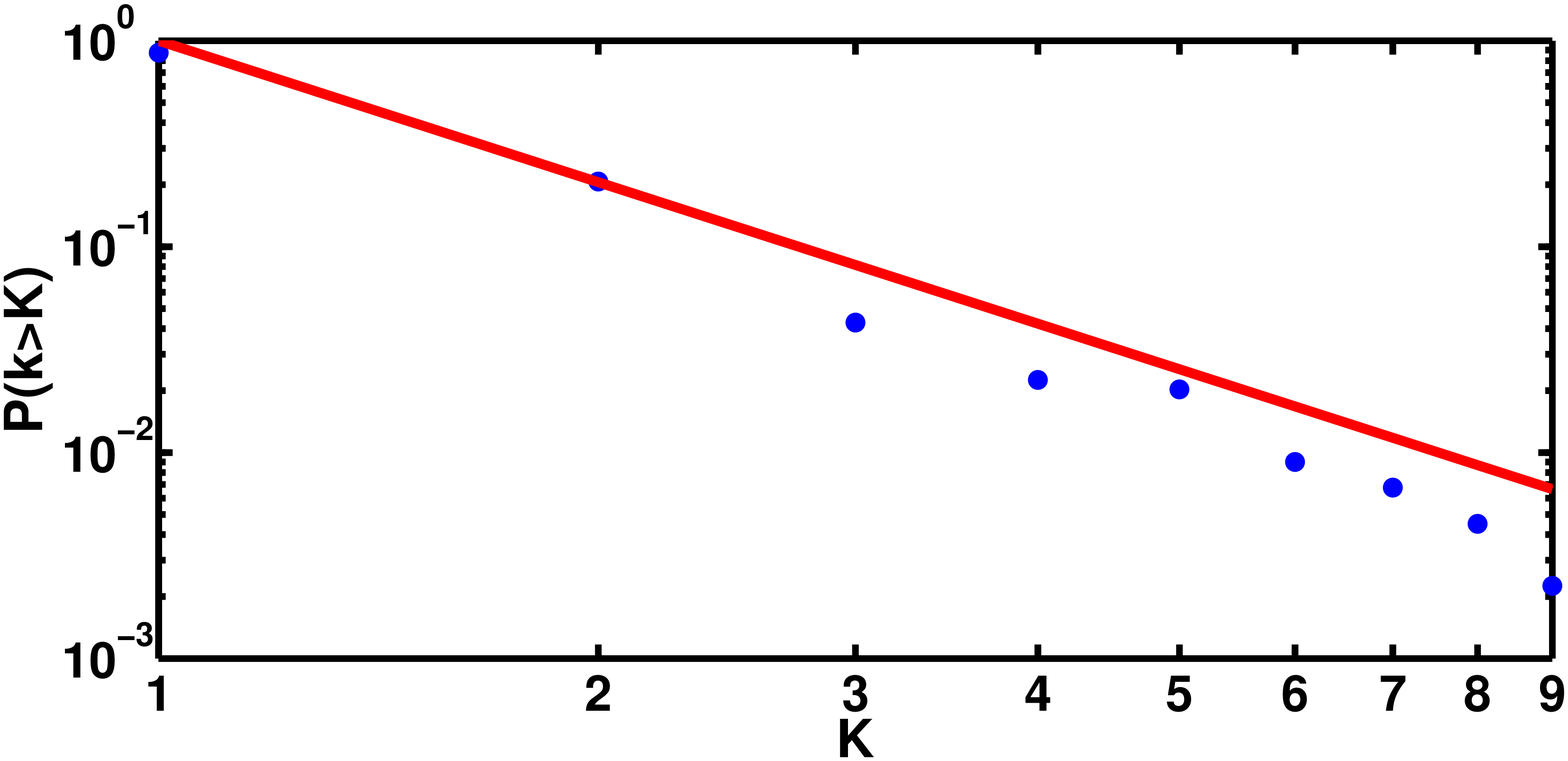}
   \caption{Node Degree Cumulative Probability Distribution for Medium Voltage sample \#3. Circles represent sample data, while straight line represents a power-law with $\gamma=2.282$.}
\label{fig:nddMV3}
\end{figure}

The most significant samples for this kind of analysis are
the biggest samples belonging to the \MV network and the most numerous ones from the
\LV (i.e., samples \#5, \#9 and \#10 from Table~\ref{fig:uLow}). All
these samples tend to follow a straight line in the log-log
plot. Figures~\ref{fig:nddLV5} and~\ref{fig:nddLV8} show the distributions
for \LV samples, while Figures~\ref{fig:nddMV2} and~\ref{fig:nddMV3}
show the probability distribution for the \MV ones.

We thus conclude that the \MV and \LV tend to be Scale-free networks,
although some exponential tail appear due to physical and economic
constraints in the network. This means robustness in terms of
redundancy of paths, but fragility to attacks on the hubs. The hubs tend to be
the few nodes that most likely lead to the High Voltage segment 
in a certain geographical location.

\subsection{Betweenness}

Betweenness describes the
importance of a node with respect to minimal paths in the graph. This
is very important to identify critical components of the \PG~\cite{Albert2000,Moreno03,Crucitti04}.
For a given node, betweenness, sometimes also referred as {\em load,} is
defined as the number of shortest paths that traverse that
node. 
\begin{definition}[Betweenness]
 The {\em betweenness} of vertex $v \in V$ is the number of shortest
 paths between any two vertices in graph $G$ that contain $v$, i.e.,
\[
 b(v)=\sum_{v}\sigma_{st}(v)
\]
where $\sigma_{st}(v)$ is the number of shortest paths from node $s$ to node $t$ traversing $v$.
\end{definition}
The betweenness is an important measure because it allows to find if
there are nodes that are critical for the whole infrastructure. In fact, the removal of nodes with the highest
betweenness can lead to critical effects on the network connectivity~\cite{Moreno03}.

\begin{figure}
   \centering
   \includegraphics[width=\textwidth]{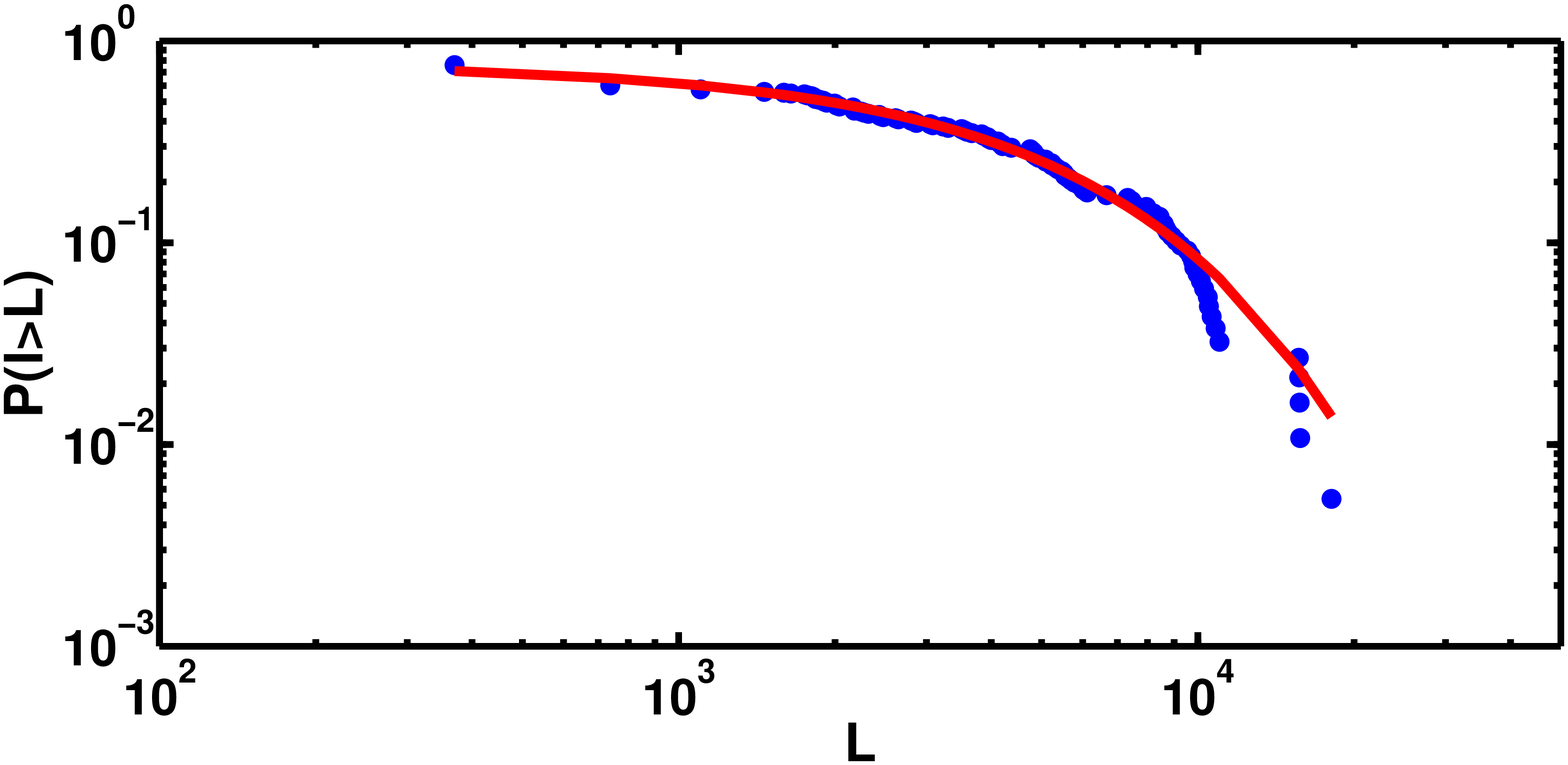}
   \caption{Betweenness Cumulative Probability Distribution for Low Voltage sample \#5 (logarithmic scale). Circles represent sample data, while continuous line represents an exponential decay $y=0.7699e^{-2.227\cdot10^{-4}x}$.}
\label{fig:bpdLV5}
\end{figure}
\begin{figure}
   \centering
   \includegraphics[width=\textwidth]{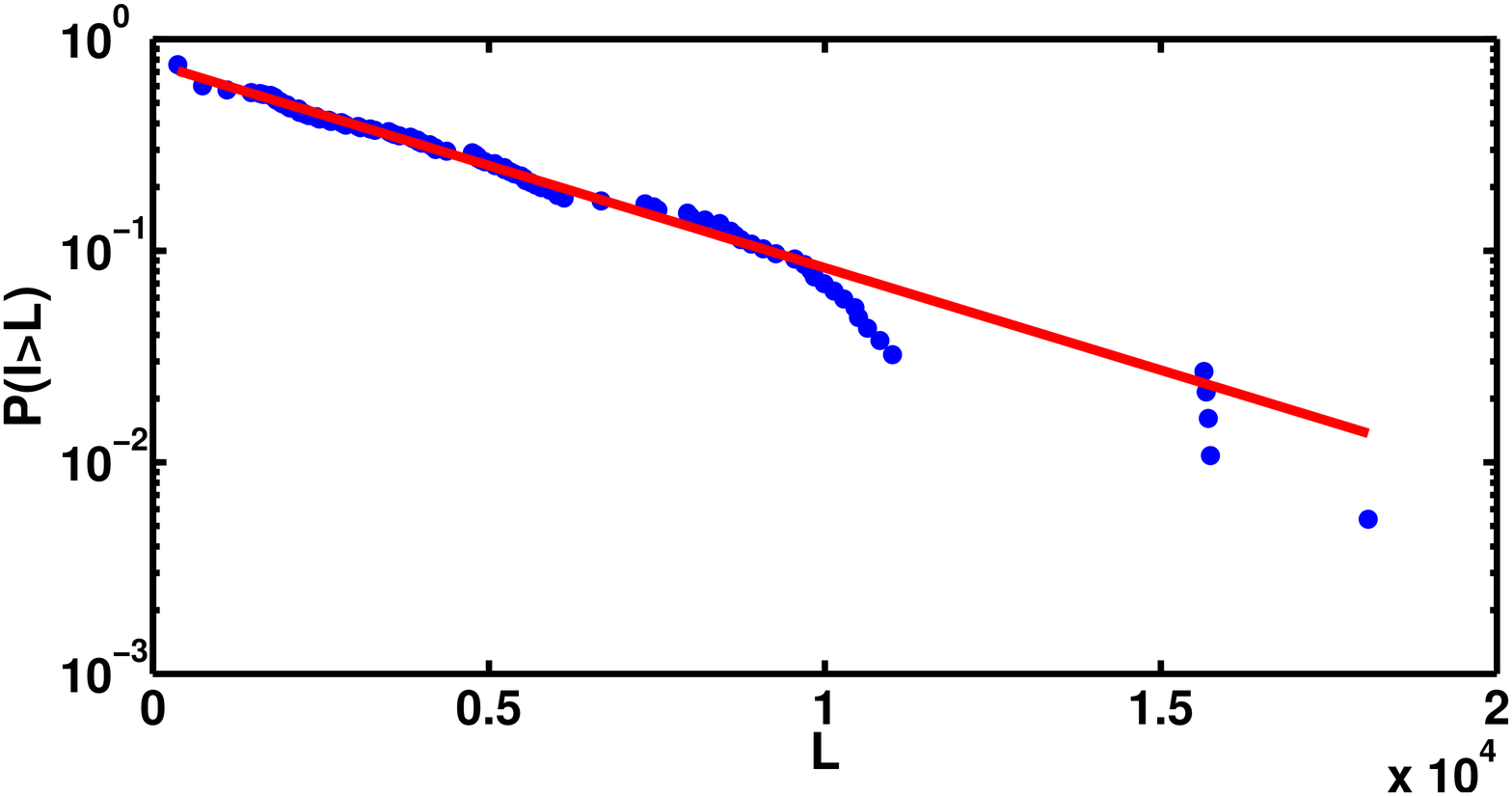}
   \caption{Betweenness Cumulative Probability Distribution for Low Voltage sample \#5 (semi-logarithmic scale). Circles represent sample data, while straight line represents an exponential decay $y=0.7699e^{-2.227\cdot10^{-4}x}$.}
\label{fig:bpdLV5semi}
\end{figure}
\begin{figure}
   \centering
   \includegraphics[width=\textwidth]{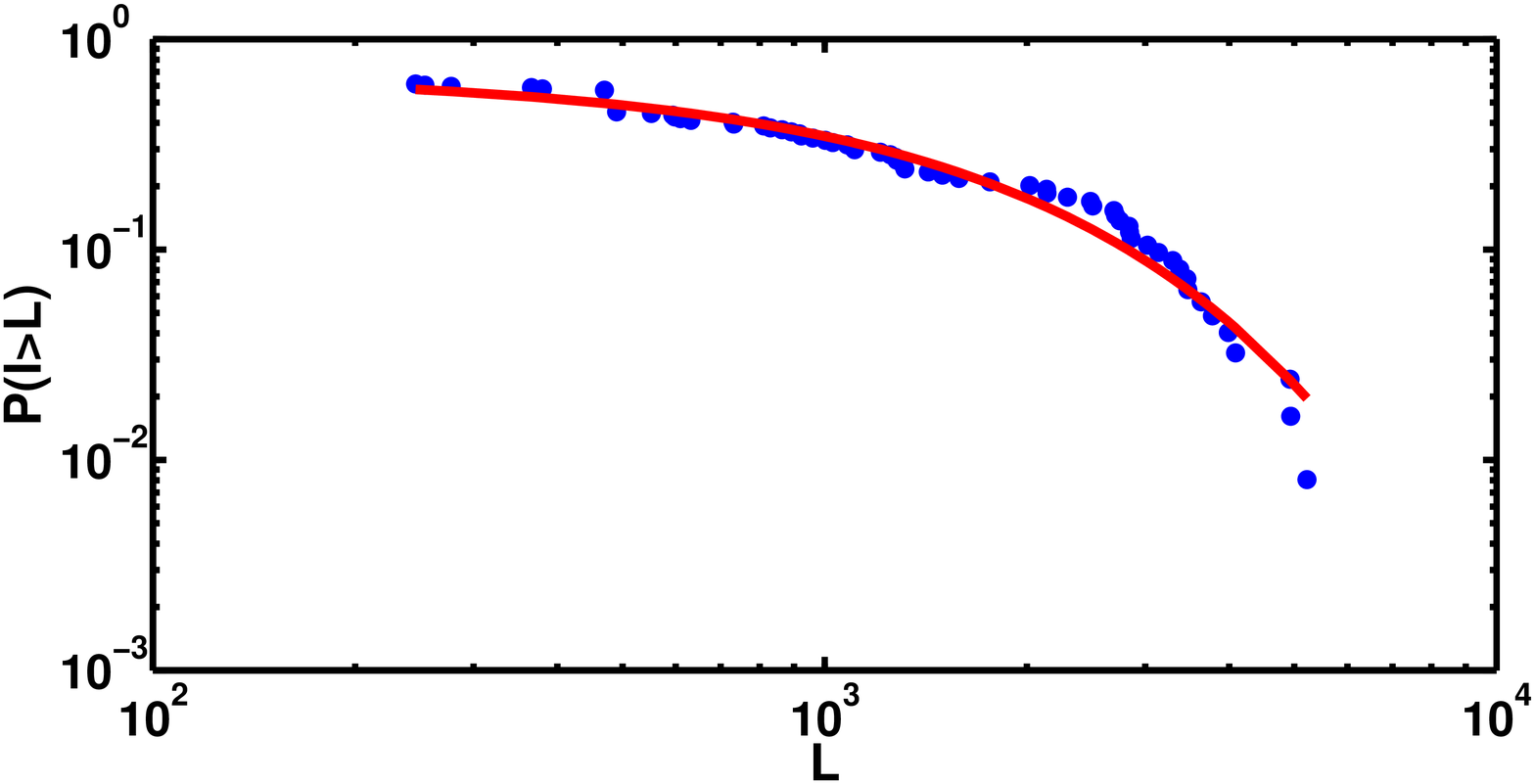}
   \caption{Betweenness Cumulative Probability Distribution for Low Voltage sample \#10 (logarithmic scale). Circles represent sample data, while continuous line represents an exponential decay $y=0.6825e^{-6.798\cdot10^{-4}x}$.}
\label{fig:bpdLV10}
\end{figure}
\begin{figure}
   \centering
   \includegraphics[width=\textwidth]{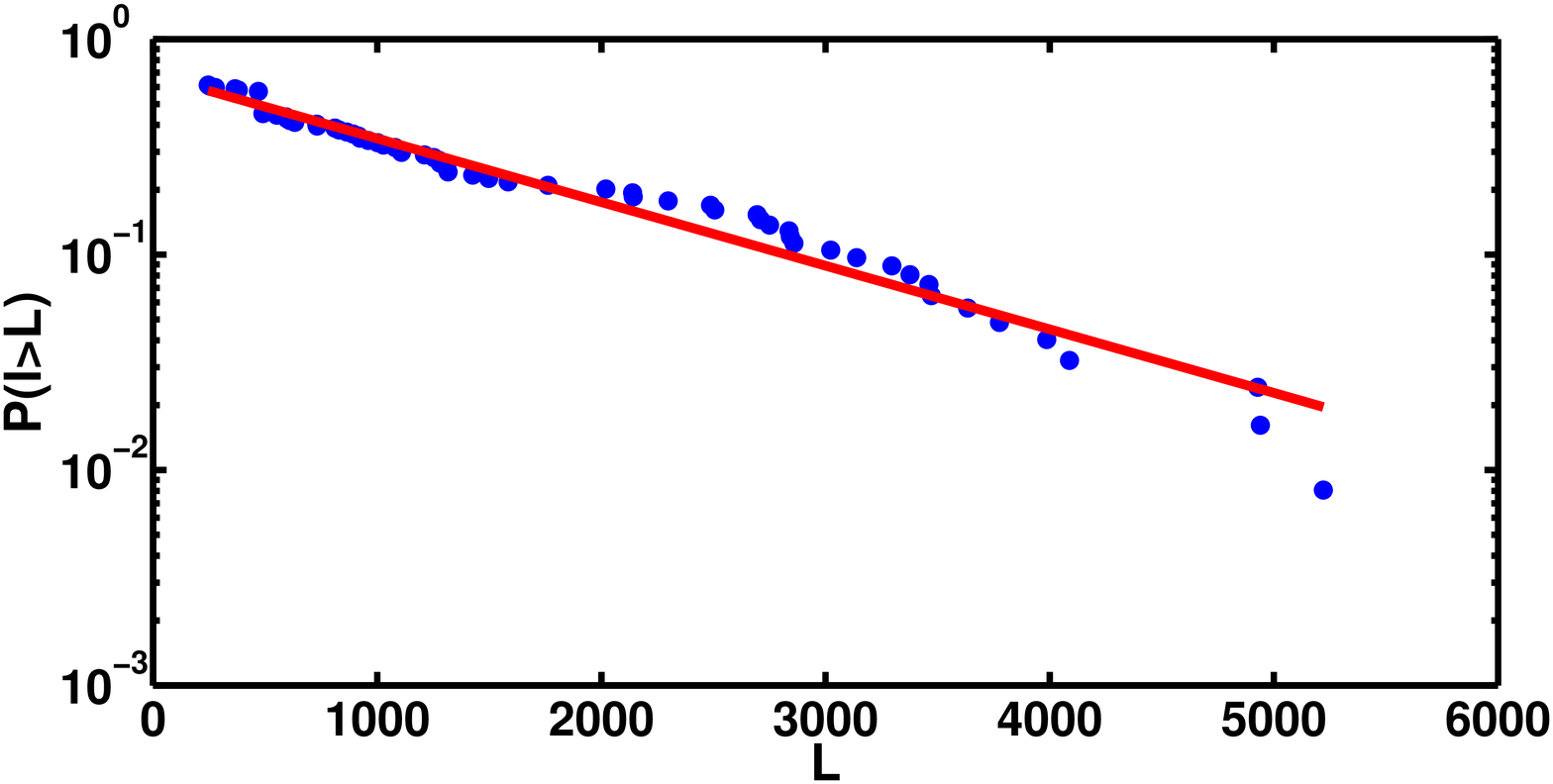}
   \caption{Betweenness Cumulative Probability Distribution for Low Voltage sample \#10 (semi-logarithmic scale). Circles represent sample data, while straight line represents an exponential decay $y=0.6825e^{-6.798\cdot10^{-4}x}$.}
\label{fig:bpdLV10semi}
\end{figure}
\begin{figure}
   \centering
   \includegraphics[width=\textwidth]{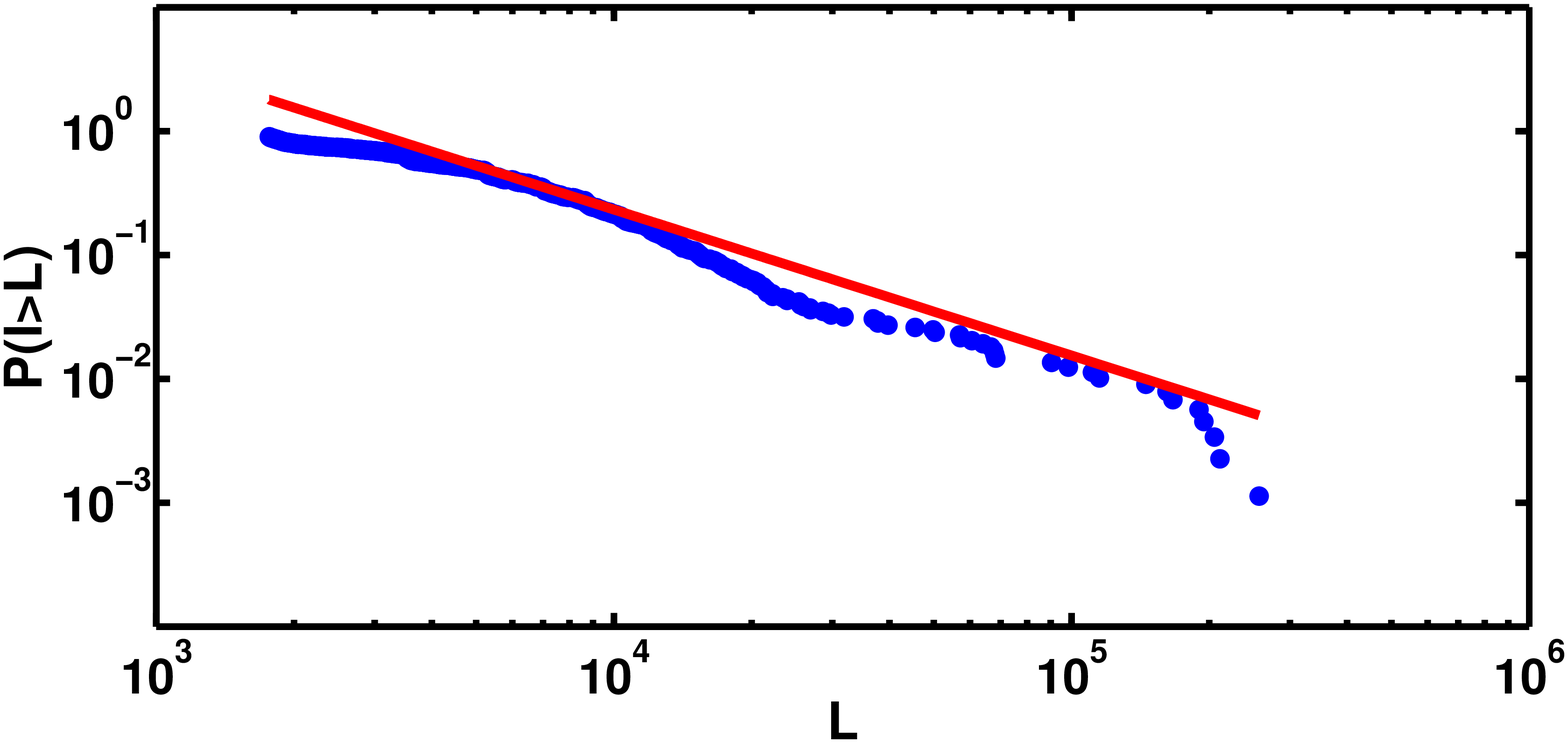}
   \caption{Betweenness Cumulative Probability Distribution for Medium Voltage sample \#2 (logarithmic scale). Circles represent sample data, while straight line represents a power-law with $\gamma=1.178$.}
\label{fig:bpdMV2}
\end{figure}
\begin{figure}
   \centering
   \includegraphics[width=\textwidth]{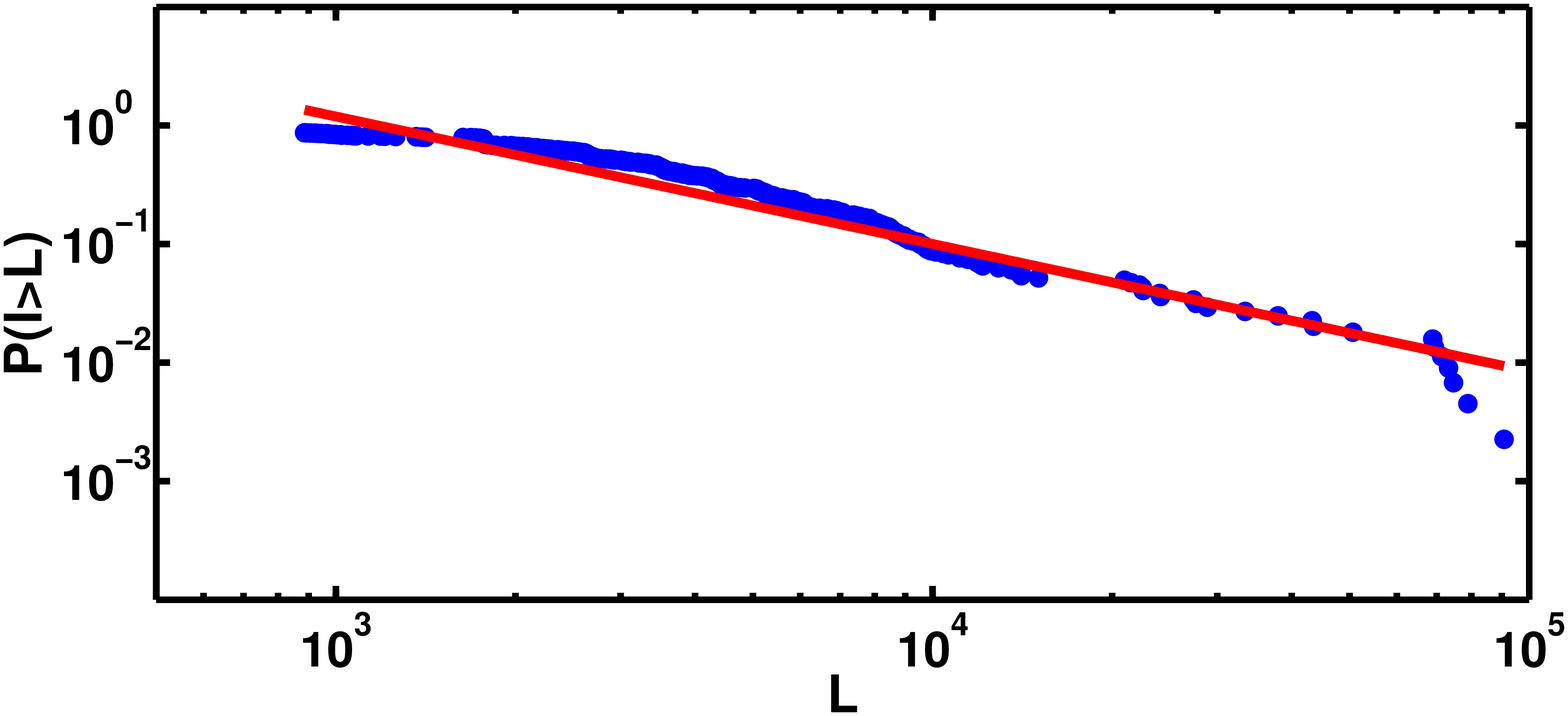}
   \caption{Betweenness Cumulative Probability Distribution for Medium Voltage sample \#3 (logarithmic scale). Circles represent sample data, while straight line represents a power-law with $\gamma=1.075$.}
\label{fig:bpdMV3}
\end{figure}

For the
most significant samples in the \LV network (i.e., samples \#5 and \#10) \bet probability distribution
follows an exponential decay, that is, the nodes with very high values of
\bet are less likely to be present in the network, as shown in Figures~\ref{fig:bpdLV5} and~\ref{fig:bpdLV10}. This aspect is not
surprising since the \LV network is quite hierarchical and the paths
tend to follow the few ones admissible by the relative simple
topology. In fact, the betweenness probability distribution charts
(Figures~\ref{fig:bpdLV5} and~\ref{fig:bpdLV10})
do not fit a 
straight line in a logarithmic plot, exhibiting a fast decay.
 This impression is also reinforced by the charts in
 Figures~\ref{fig:bpdLV5semi} and~\ref{fig:bpdLV10semi}, that is, the same
betweenness probability distribution, but on a log-linear scale: the
straight line in this kind of diagram is a sign of an exponential
decay,~\cite{Amaral2000}. 
In addition a fitting procedure, using the
non-linear least square method gives very good
results approximating the \bet probability distribution samples with
an exponential function. On the other hand, \bet in \MV segment seems
to follow a power-law decay, this is shown in the logarithmic chart in
Figures~\ref{fig:bpdMV2} and~\ref{fig:bpdMV3}.
The samples from \MV
network show a distribution of \bet with a much fatter tail than the
\LV ones, that is there are nodes that are central in many paths. This
is due to the more meshed structure the \MV network has, compared to the
\LV one. This result for \MV \bet is closer to the results obtained
for this same metric in \HV studies,~\cite{Albert2000,Crucitti04}. In summary, a few nodes
are extremely critical to enable the electricity distribution to
the whole network.

\subsection{Node importance}
\label{sec:unwRes}
To have a general understanding of the critical elements of the
Grid, we resort to the matrix representation of the
graph and study the eigenvalues and eigenvectors to find the most
important and critical nodes of the network.
\begin{definition}[Adjacency matrix]
The adjacency matrix $A=A(G)=(a_{i,j})$ of a graph $G$ is the $n \times n$ matrix given by
\[
 a_{ij} =
  \begin{cases}
   1      & \text{if } e_{i,j}=(v_i,v_j) \in E, \\
   0       & \text{otherwise.}
  \end{cases}
\]
\end{definition}
 {\em Centrality} refers
to the importance of the node in terms of degree, betweenness,
closeness or eigenvectors~\cite{newman08,BONACICH2007}. In this work,
we use eigenvector centrality to stress the dependence of the
centrality of one node with the centrality of the other nodes it is
connected to. The components of the dominant
eigenvector then represent the nodes of the graph. The highest the value, the highest
the centrality of that node, the highest the importance of that node
in the Grid. 

The most critical nodes for the most
interesting samples from \LV are shown in Tables~\ref{tab:eigvcentral5LV} and~\ref{tab:eigvcentral10LV} while the results for
samples from \MV are shown in Tables~\ref{tab:eigvcentral2MV} and~\ref{tab:eigvcentral3MV}. The first column of each table represents the ranking position for the first ten node whose identifier is given in the second column.
As mentioned above there are several measure to identify the most
important nodes in a network. The core aspect of eigenvector
centrality is that the importance of a node is dependent from the
importance of their neighbouring nodes, and therefore to some extent
from all nodes in a connected network. This is not the case for other
measures of centrality (e.g., node degree centrality and betweenness
centrality) whose values are not influenced by the properties of other
nodes.

\begin{table}[h!b!p!]
\begin{center}
\begin{footnotesize}
    \begin{tabular}{ | p{3cm} | p{3cm}|}
    \hline
    Eigenvector Centrality Ranking \# & Node ID \\ \hline
1&10\\ \hline
2&93\\ \hline
3&111\\ \hline
4&148\\ \hline
5&transformer 5\\ \hline
6&22\\ \hline
7&28\\ \hline
8&27\\ \hline
9&26\\ \hline
10&25\\ \hline
    \end{tabular}
    \end{footnotesize}
\end{center}
\caption{Eigenvector centrality ranking for \LV sample \#5.}
\label{tab:eigvcentral5LV}
\end{table}

\begin{table}[h!b!p!]
\begin{center}
\begin{footnotesize}
    \begin{tabular}{ | p{3cm} | p{3cm}|}
    \hline
    Eigenvector Centrality Ranking \# & Node ID \\ \hline
1&3	\\ \hline
2&44\\ \hline
3&108\\ \hline
4&39\\ \hline
5&109\\ \hline
6&110\\ \hline
7&102\\ \hline
8&107\\ \hline
9&2\\ \hline
10&61\\ \hline
    \end{tabular}
    \end{footnotesize}
\end{center}
\caption{Eigenvector centrality ranking for \LV sample \#10.}
\label{tab:eigvcentral10LV}
\end{table}

\begin{table}[h!b!p!]
\begin{center}
\begin{footnotesize}
    \begin{tabular}{ | p{3cm} | p{3cm}|}
    \hline
    Eigenvector Centrality Ranking \# & Node ID\\ \hline
1&546\\ \hline
2&574\\ \hline
3&608\\ \hline
4&609\\ \hline
5&582\\ \hline
6&32\\ \hline
7&580\\ \hline
8&56\\ \hline
9&9\\ \hline
10&765\\ \hline
    \end{tabular}
    \end{footnotesize}
\end{center}
\caption{Eigenvector centrality ranking for \MV sample \#2.}
\label{tab:eigvcentral2MV}
\end{table}

\begin{table}[h!b!p!]
\begin{center}
\begin{footnotesize}
    \begin{tabular}{ | p{3cm} | p{3cm}|}
    \hline
    Eigenvector Centrality Ranking \# & Node ID\\ \hline
1&351\\ \hline
2&263\\ \hline
3&324\\ \hline
4&6\\ \hline
5&12\\ \hline
6&350\\ \hline
7&299\\ \hline
8&11\\ \hline
9&80\\ \hline
10&355\\ \hline
    \end{tabular}
    \end{footnotesize}
\end{center}
\caption{Eigenvector centrality ranking for \MV sample \#3.}
\label{tab:eigvcentral3MV}
\end{table}

\subsection{Fault tolerance}

A related study is to evaluate the reliability of a network by
analysing its connectivity when nodes are removed. There are basically
two ways to perform this analysis: choosing the nodes randomly or
selecting the nodes to be removed following a certain property or
metric significant for the network. Similar studies concerning the
resilience of the \HV \PG exist, e.g.,~\cite{casals07,Albert2000}. We 
apply such technique to the \MV and \LV ends of the Power Grid using three
policies for node removal: random, highest degree
and highest betweenness driven choices. The measure that is taken into
account is the order of the largest connected component of the
network (i.e., the number of nodes composing the biggest cluster in the network) computed as a fraction of the original order of the
network, and its evolution while nodes of the network are removed,
again the latter are considered as a fraction of the original order
of the network.

\begin{figure}
   \centering
   \includegraphics[width=\textwidth]{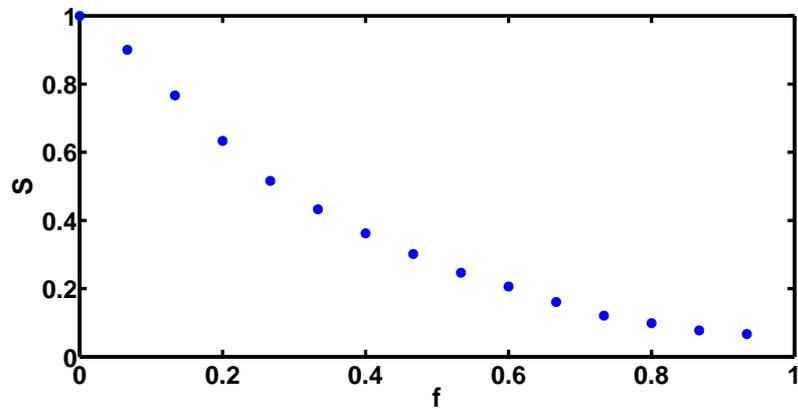}
   \caption{Resilience for random-based removal for \LV sample \#2. The horizontal axis represents the fraction $f$ of the nodes removed from the original sample; the vertical axis represents the size of the largest connected component $S$ relative to the initial size of the graph.}
\label{fig:remRANLV2}
\end{figure}

\begin{figure}
   \centering
   \includegraphics[width=\textwidth]{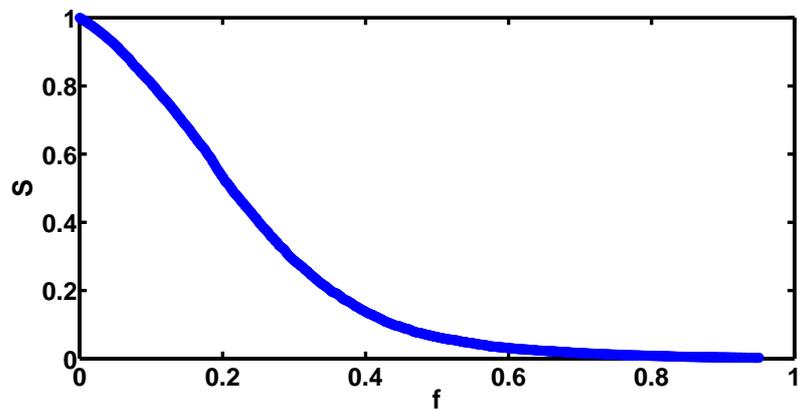}
   \caption{Resilience for node random-based removal for \MV sample \#2. The horizontal axis represents the fraction $f$ of the nodes removed from the original sample; the vertical axis represents the size of the largest connected component $S$ relative to the initial size of the graph.}
\label{fig:remRANMV12}
\end{figure}

The {\em random removal} simulates casual errors. 
As shown in~\cite{Cohen2000}, networks
that follow a power-law whose characteristic parameter $\gamma < 3$
tend to have a high value for the transition threshold at which they
disrupt.
In the samples analysed it seems that this is true
especially for the small samples that generally have
a cluster that is 10\% of the original when almost 90\% of the nodes
are removed, as shown in Figure~\ref{fig:remRANLV2}. The situation is different for samples with higher order
that show a cluster that is
reduced to 10\% of the original when about 40\% of the nodes are
removed, as shown in Figure~\ref{fig:remRANMV12}. Even if the degree distributions found for samples following a power-law have a parameter $\gamma < 3$
the samples show a threshold effect that is more similar, according to~\cite{Cohen2000}, to networks whose characteristic $\gamma > 3$.

\begin{figure}
   \centering
   \includegraphics[width=\textwidth]{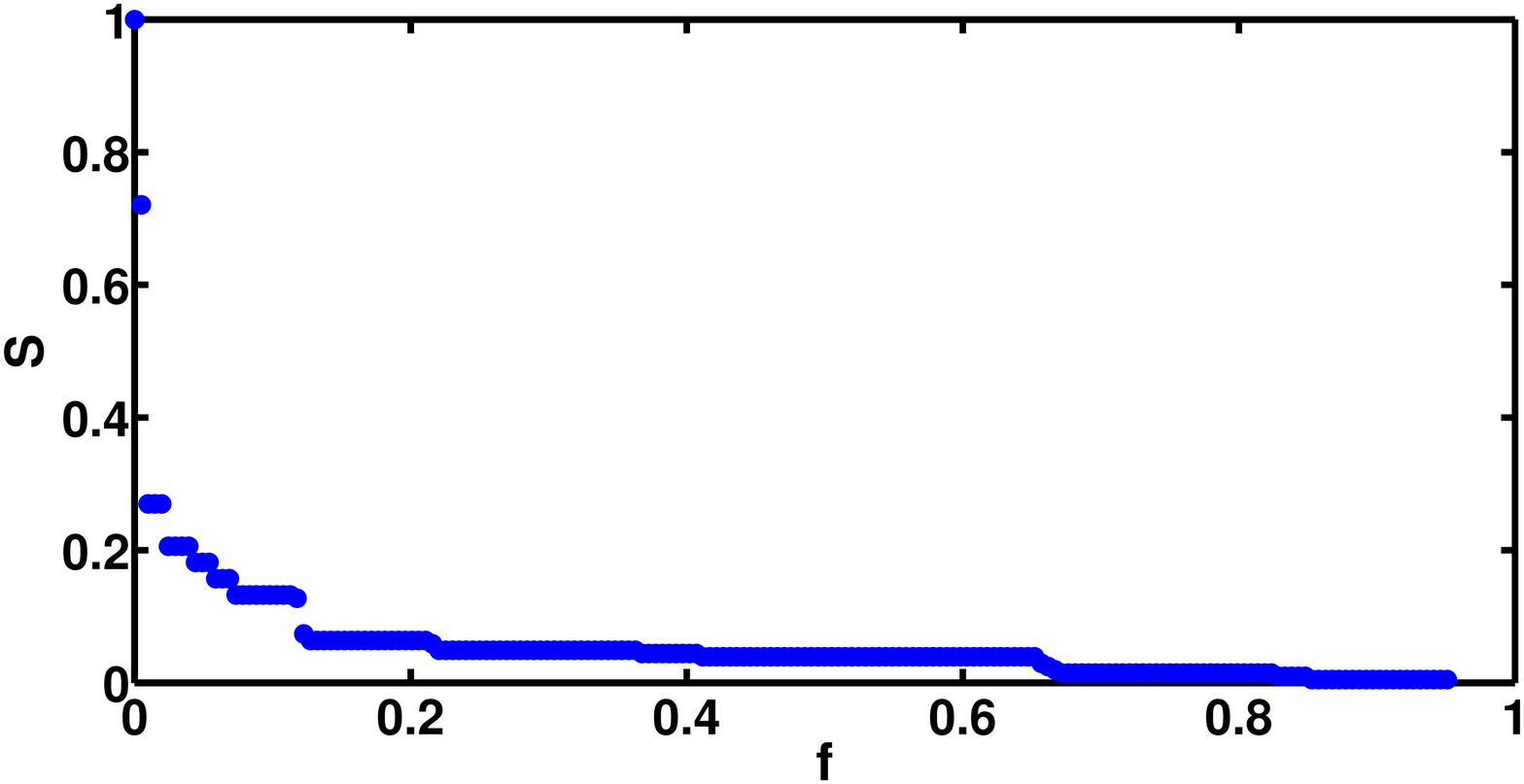}
   \caption{Resilience for node degree-based removal for \MV sample \#10. The horizontal axis represents the fraction $f$ of the nodes removed from the original sample; the vertical axis represents the size of the largest connected component $S$ relative to the initial size of the graph.}
\label{fig:remNDMV10}
\end{figure}

\begin{figure}
   \centering
   \includegraphics[width=\textwidth]{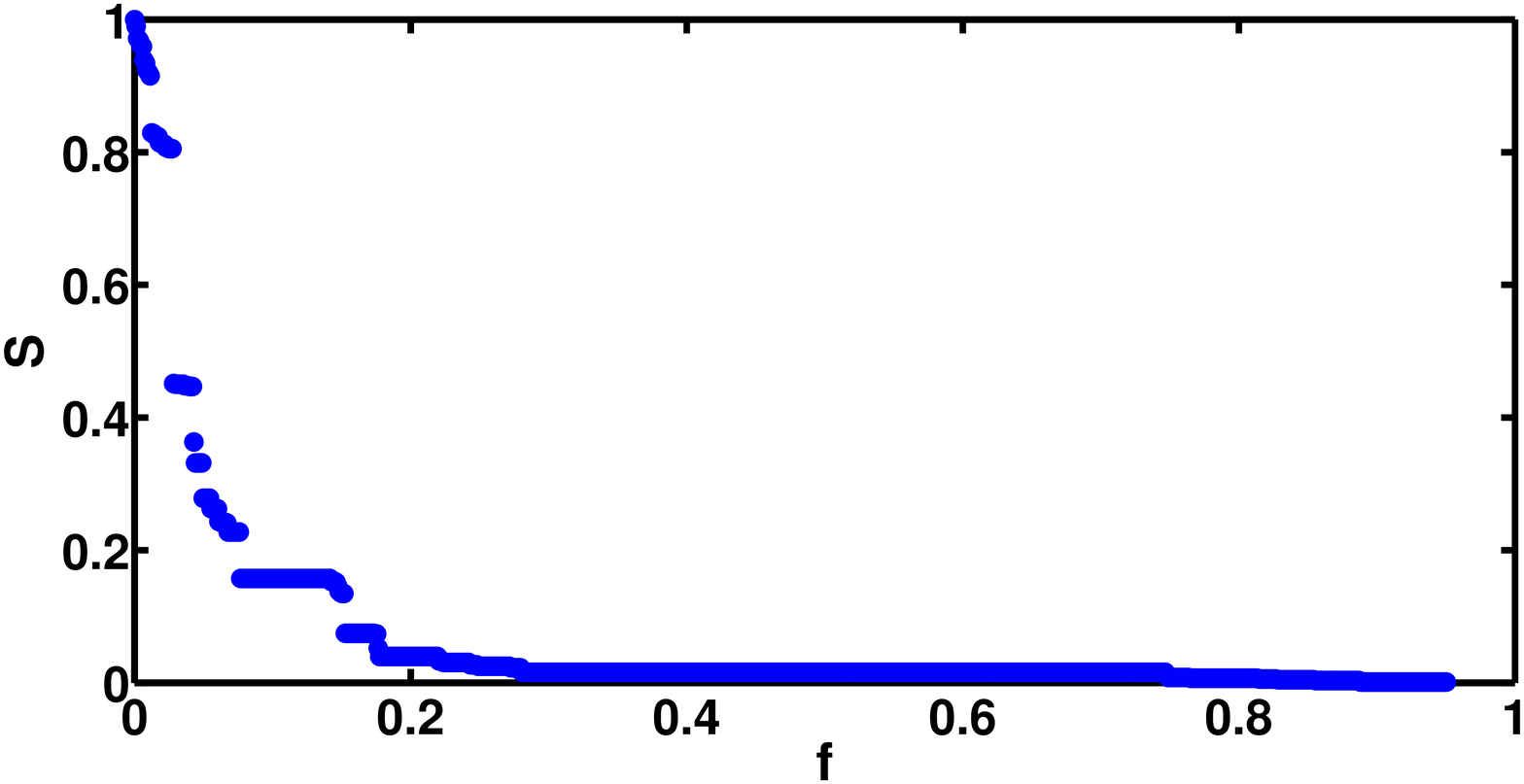}
   \caption{Resilience for node degree-based removal for \MV sample \#2. The horizontal axis represents the fraction $f$ of the nodes removed from the original sample; the vertical axis represents the size of the largest connected component $S$ relative to the initial size of the graph.}
\label{fig:remNDMV2}
\end{figure}

\begin{figure}
   \centering
   \includegraphics[width=\textwidth]{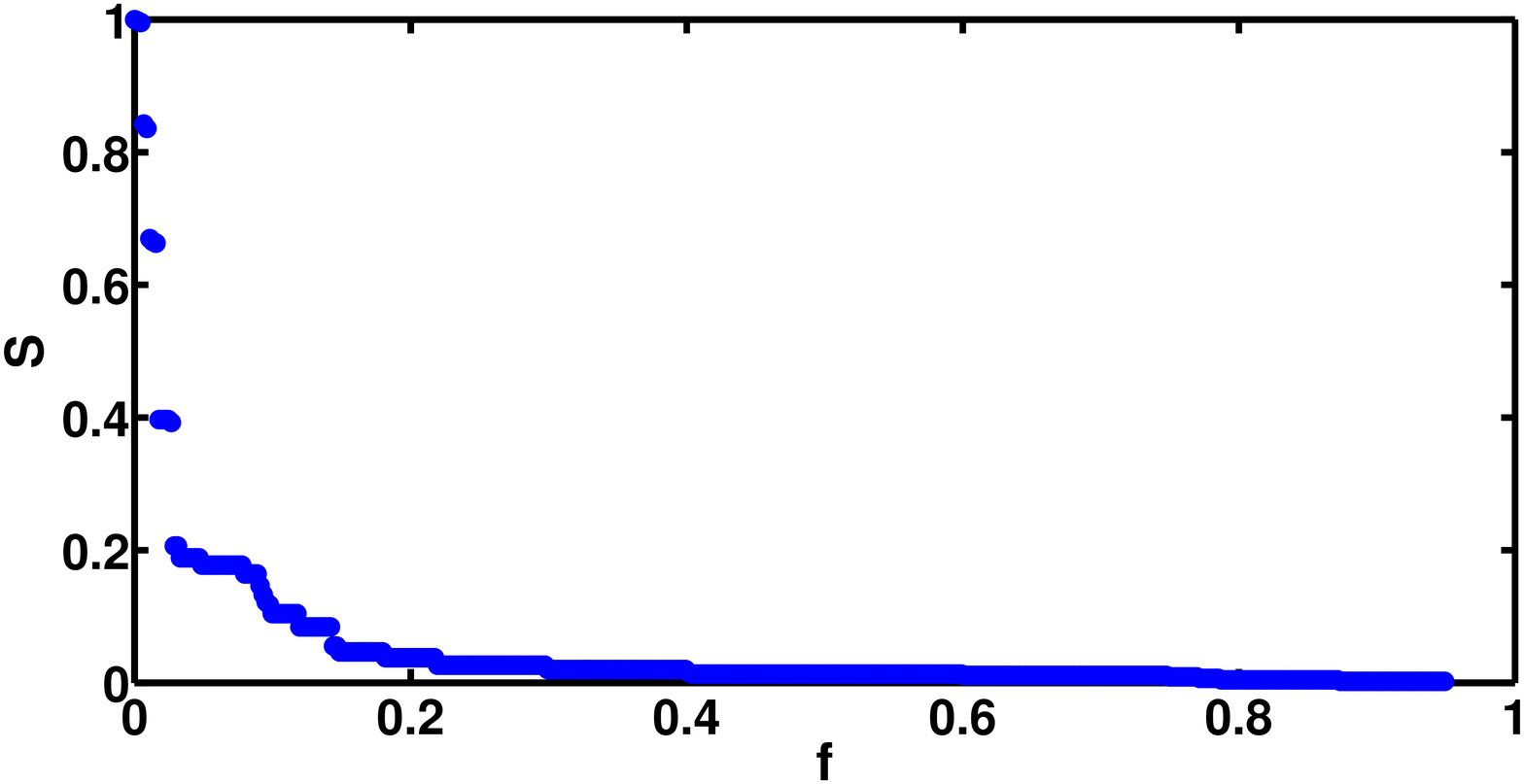}
   \caption{Resilience for node degree-based removal for \MV sample \#3. The horizontal axis represents the fraction $f$ of the nodes removed from the original sample; the vertical axis represents the size of the largest connected component $S$ relative to the initial size of the graph.}
\label{fig:remNDMV3}
\end{figure}

\begin{figure}
   \centering
   \includegraphics[width=\textwidth]{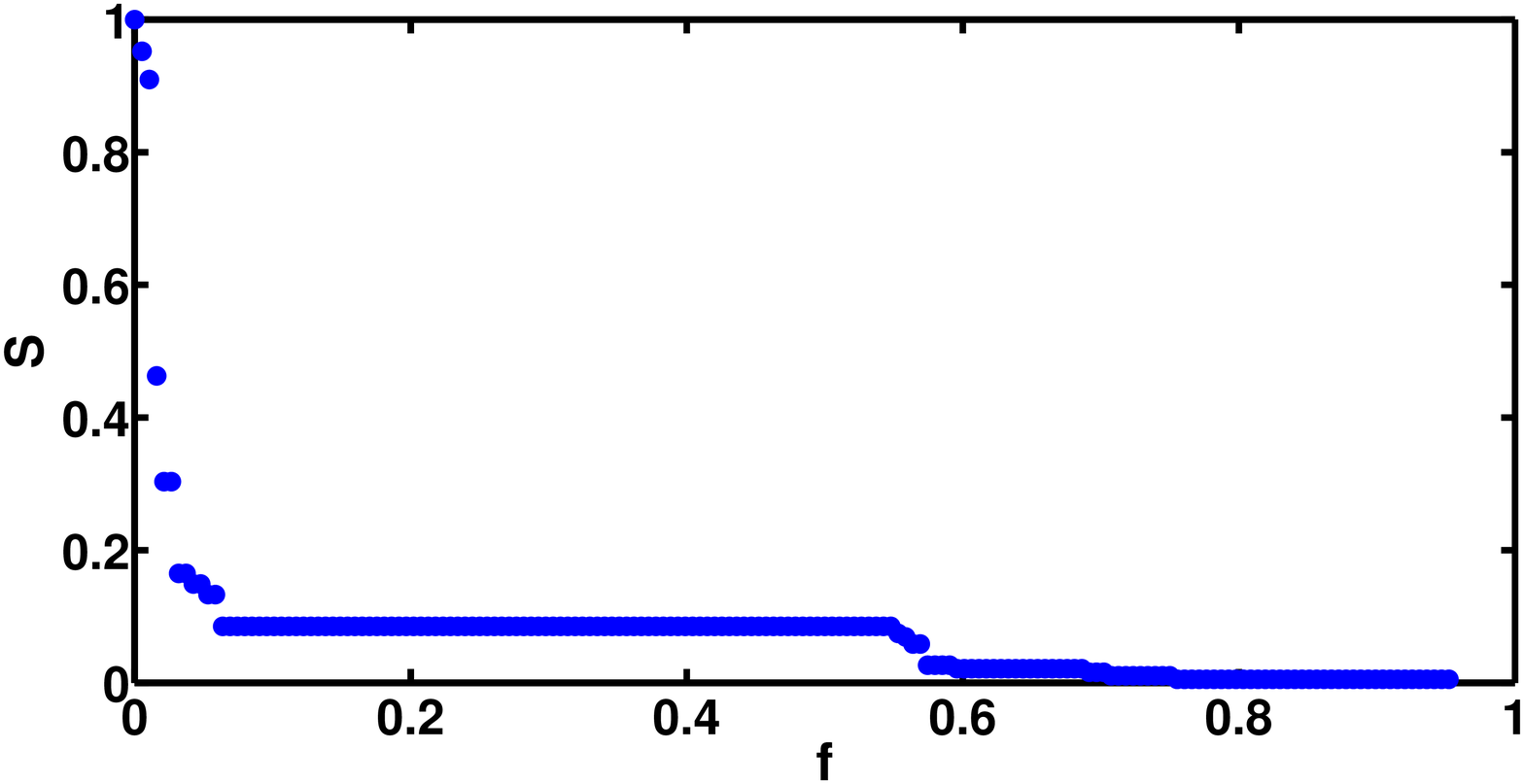}
   \caption{Resilience for node degree-based removal for \LV sample \#5. The horizontal axis represents the fraction $f$ of the nodes removed from the original sample; the vertical axis represents the size of the largest connected component $S$ relative to the initial size of the graph.}
\label{fig:remNDLV5}
\end{figure}

\begin{figure}
   \centering
   \includegraphics[width=\textwidth]{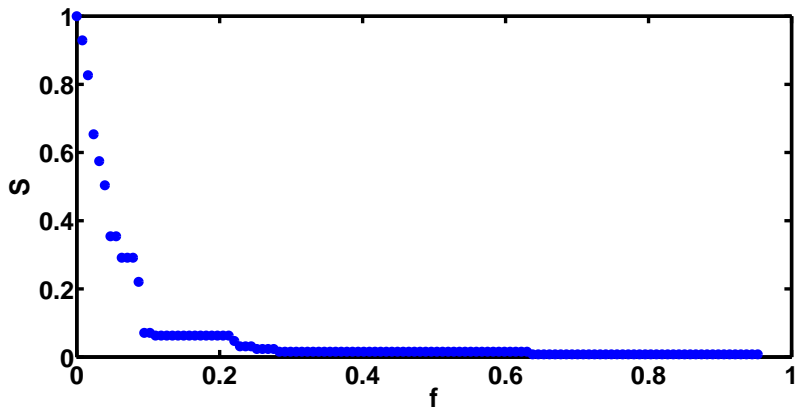}
   \caption{Resilience for node degree-based removal for \LV sample \#10. The horizontal axis represents the fraction $f$ of the nodes removed from the original sample; the vertical axis represents the size of the largest connected component $S$ relative to the initial size of the graph.}
\label{fig:remNDLV10}
\end{figure}

The situation is radically different when ``targeted attacks'' are
considered. In particular two kind of attack policies are
investigated: {\em node degree-based removal} and {\em betweenness-based
removal.} The main difference compared to the random-based removal
is the presence of very sharp falls that appear when certain nodes are
targeted. The removal of selected nodes can
cause a drop in the size of the maximal connected component even of
40\%, as shown in Figure~\ref{fig:remNDMV10}. Node degree-based
removal is much more critical than the random removal: by just
removing 10\% of the most connected
nodes one reduces the network to only 10\% of its original size. The same
applies for the biggest samples considered both in the \LV and \MV
network, as shown in Figures~\ref{fig:remNDLV5}
and~\ref{fig:remNDLV10} for the \LV and Figure~\ref{fig:remNDMV2}
and~\ref{fig:remNDMV3} for the Medium voltage.


\begin{figure}
   \centering
   \includegraphics[width=\textwidth]{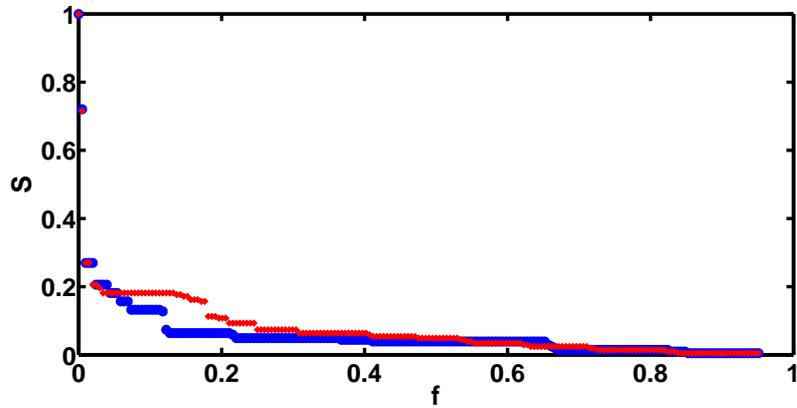}
   \caption{Resilience for node degree-based and betweenness-based removal for \MV sample \#10. The horizontal axis represents the fraction $f$ of the nodes removed from the original sample; the vertical axis represents the size of the largest connected component $S$ relative to the initial size of the graph. Red diamonds represent the betweenness-based removal, while blue circles represent the node node degree-based.}
\label{fig:remcombMV10}
\end{figure}

\begin{figure}
   \centering
   \includegraphics[width=\textwidth]{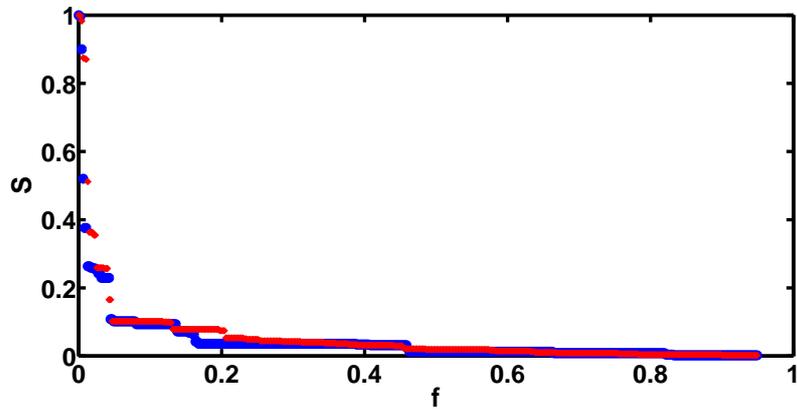}
   \caption{Resilience for node degree-based and betweenness-based removal for \MV sample \#13. The horizontal axis represents the fraction $f$ of the nodes removed from the original sample; the vertical axis represents the size of the largest connected component $S$ relative to the initial size of the graph. Red diamonds represent the betweenness-based removal, while blue circles represent the node node degree-based.}
\label{fig:remcombMV13}
\end{figure}

The removal of nodes based on the highest betweenness shows generally
the same behaviour, as degree-based removal, with network disruption
that appear much faster than random-based network failures. Considering
the general correlation between nodes with a certain degree and their
betweenness it is not surprising that the two removal policies have
very similar results and shape. The only remark that generally
differentiates the betweenness-based removal is a little higher order
of the maximal connected component compared to the one obtained with a
degree-based removal when the same fraction of nodes is removed. In
addition the decrease of the order of the maximal connected component
tends to be slightly smoother than the degree-based one. Figures~\ref{fig:remcombMV10} and~\ref{fig:remcombMV13} show the comparison
of the two removal policies for the samples that show some interesting
deviations in the correlation of the degree and betweenness.


Another way to investigate the resilience is 
to determine which and how many edges have to be removed to break the graph into two disconnected components almost of the same order. 
This information is essential to have a general clue of the robustness of the Grid, that is how many transmission lines must be broken simultaneously to split the network in two components with almost the same amount of nodes,~\cite{Rosato2007}. There are several methods and algorithms to solve this kind of problem that can be interpreted as a classic \emph{max-flow/min-cut} problem, however an interesting method exploits the spectrum of the Laplacian matrix of a graph. 
\begin{definition}[Laplacian matrix]
Let $D=(D_{ij})$ be a diagonal matrix with $D_{ii}=d(v_i)$ the degree of vertex $v_i$ in graph $G$ and $A$ the adjacency matrix of $G$.\\
The matrix $L=D-A$ is the Laplacian matrix of graph $G$.
\end{definition}
In particular this method exploits the second smallest eigenvalue of the Laplacian matrix of the corresponding graph and computes the corresponding eigenvector (for a complete explanation of eigenvalues of Laplacian matrix spectral properties we refer to ~\cite{Mohar1992,Mohar1991}). This resulting eigenvector has components, each one representing a node of the graph, that can be either positive or negative. Each node whose eigenvector component is positive belongs to one sub-graph, while the ones with negative components belong to the other sub-graph. The edges that connect nodes belonging to the different sub-graphs are the critical edges that if removed lead to two different sub-graphs. The more edges connect the two sub-graphs, the more robust is the Grid. Once the two sub-graphs are identified it is possible to iterate the method on each sub-graph and find again the most critical edges.

The number of critical edges that, if removed at the same time, disrupt the network evenly in two or more sub-networks, are shown in Table \ref{tab:eigvcritLV} and Table \ref{tab:eigvcritMV} for \LV and \MV respectively. The first column represent the sample identifier, while the second column represent the number of edges to be removed simultaneously to split evenly the network. Generally \MV networks are more robust to edge failures than \LV with the number of edges to be removed that is double. \MV sample \#2 is the most reliable to line disruption attack and indeed this is an indication of the high mashed structure the biggest sample analysed owns.

\begin{table}[h!b!p!]
\begin{center}
\begin{footnotesize}
    \begin{tabular}{ | p{3cm} | p{3cm}| p{3cm}|}
    \hline
    Sample \# & Number of Critical Edges\\ \hline
1&2\\ \hline
2&2\\ \hline
3&2\\ \hline
4&1\\ \hline
5&2\\ \hline
6&2\\ \hline
7&1\\ \hline
8&1\\ \hline
9&2\\ \hline
10&3\\ \hline
11&1\\ \hline
    \end{tabular}
    \end{footnotesize}
\caption{Number of critical edges according to Laplacian eigenvalue method for \LV samples.}
\label{tab:eigvcritLV}
\end{center}
\end{table}

\begin{table}[h!b!p!]
\begin{center}
\begin{footnotesize}
    \begin{tabular}{ | p{3cm} | p{3cm}| p{3cm}|}
    \hline
    Sample \# & Number of Critical Edges\\ \hline
1&2\\ \hline
2&27\\ \hline
3&4\\ \hline
4&5\\ \hline
5&3\\ \hline
6&4\\ \hline
7&4\\ \hline
8&1\\ \hline
9&6\\ \hline
10&4\\ \hline
11&4\\ \hline
12&1\\ \hline
13&6\\ \hline

    \end{tabular}
    \end{footnotesize}
\caption{Number of critical edges according to Laplacian eigenvalue method for \MV samples.}
\label{tab:eigvcritMV}
\end{center}
\end{table}

In summary, the results for the \LV and \MV show disruption behaviours. These networks are quite immune to random
failures to which the networks present a constant degrading
disruption, while they deeply suffer from certain characteristic nodes to be
removed.

\section{Weighted \PG study}\label{sec:weighted}

The purely topological study of the \PG just presented already
gives important information about the connectivity and robustness of the
Medium and Low Voltage Grids, though it does not consider the
different physical properties of the cables. These can vary greatly
for different sections of the Grid and provide essential indications
to establish the behaviour of a link. Next we perform an analysis of
the same samples of the Grid also considering the 
resistance as the weight of the graph model of 
Definition~\ref{def:wpgg}.

\begin{table}[htb]
\begin{center}
\begin{footnotesize}
    \begin{tabular}{ | p{0.5cm} ||p{2cm}| p{2cm} |p{2cm} |}
    \hline
    ID & Weighted Characteristic Path Length& Edge Average Weight& Normalized Weighted Characteristic Path Length\\ \hline\hline
1&2.000	&0.698	&	2.865\\ \hline
2&1.429	&0.595	&	2.402\\ \hline
3&3.066	&0.739	&	4.149\\ \hline
4&3.087	&0.699	&	4.414\\ \hline
5&12.136&	0.741	&	16.378\\ \hline
6&3.889	&1.648	&	2.360\\ \hline
7&4.162	&0.348	&	11.960\\ \hline
8&5.112	&0.876	&	5.836\\ \hline
9&7.872	&0.583	&	13.503\\ \hline
10&6.407&	0.785	&	8.162\\ \hline
11&2.967&	0.592	&	5.012\\ \hline
    \end{tabular}
\end{footnotesize}
\end{center}
 \caption{Weighted analysis of the Low Voltage samples from the northern Netherlands Power Grid.
\label{tab:weightedlow}}
\end{table}

Take the samples analyzed in Tables~\ref{fig:uLow}
and~\ref{fig:uMed}, but now consider the weighted graph
definition. The notion of a characteristic path length can be extended
to take the weights into account yielding the values shown in
Tables~\ref{tab:weightedlow} and~\ref{tab:weightedmedium}. In each
table, the second column contains the characteristic path length
resulting in the weighted graph (WCPL), formally:
\begin{definition}[Weighted characteristic path length (WCPL)]
The
\emph{weighted characteristic path length} for graph $G$, $L_{wcpl}$ is the
median for all $(v_i,v_j) \in V$ of the following distance
\[
 d_w(v_i,v_j) =  \sum_{e_{s,t}} e_{w_{s,t}}
\]
such that $e_{w_{s,t}}$ is an edge in the minimal weighted path between
$v_i$ and $v_j$.
\end{definition}
The third column provides the average value of the weights of all edges; while the fourth column shows a normalized value for the weighted characteristic path length (NWCPL) obtained by dividing the WCPL by the
average weight of the edge belonging to the same data sample. This normalization is performed to have a measure to compare the unweighted and the weighted samples whose results are shown in Section~\ref{sec:unweightedweightedcomparison}.

\begin{table*}[htb]
\begin{center}
\begin{footnotesize}
    \begin{tabular}{ | p{0.5cm} || p{2cm}| p{2cm} |p{2cm}  |}
    \hline
    ID & Weighted Characteristic Path Length & Edge Average Weight & Normalized Weighted Characteristic Path Length\\ \hline\hline
1&185.916&	12.779	&	14.549\\ \hline
2&108.011&	11.851	&	9.987\\ \hline
3&153.402&	8.608	&	17.821\\ \hline
4&163.067&	9.217	&	17.692\\ \hline
5&127.258&	7.122	&	17.868\\ \hline
6&134.661&	13.106	&	10.275\\ \hline
7&187.084&	16.382	&	11.420\\ \hline
8&148.058&	7.193	&	20.584\\ \hline
9&99.385&	7.421	&	13.392\\ \hline
10&126.845&	6.850	&	18.518\\ \hline
11&92.060&	8.764	&	10.504\\ \hline
12&38.084&	6.915	&	5.507\\ \hline
13&232.475&	13.810&		16.834\\ \hline
    \end{tabular}
\end{footnotesize}
\end{center}
 \caption{Weighted analysis of the Medium Voltage samples from the northern Netherlands Power Grid.
\label{tab:weightedmedium}}
\end{table*}

Due to the relative short length of the \LV networks cables, the WCPLs
for this segment of the network are small, as well as the average
weight of each edge (almost all of them are below the unit). The
situation is different for the \MV networks which are higher since the
cables and paths span across wider geographical areas. 
The discrepancy can be explained by the different purpose for
which these networks are designed: a bridge network from \HV transmission lines and end-user distribution (\MV network) and the final end delivery (\LV network). In fact, both the WCPL and the edge average
weight for \MV samples are approximately two order of magnitude greater than the \LV
ones. This is indeed due to an extension of \MV cables that range from hundred meters to kilometres, while \LV extend usually around tens of metres.

\subsection{Weighted Node Degree Distribution}

Though no value is associated to a node, the weights of the incident
edges also influence the node properties. One way of seeing this, is by
defining a weighted node degree.
\begin{definition}[Weighted degree]
Let $x \in V$ be a vertex in a weighted graph $G$, the \emph{weighted degree} of $x$, $d_w(x)$ is:
\[
 d_w{(x)} = \sum_{y \in \Gamma(x)}w_{x,y}
\]
where $w_{x,y}$ is the weight of the edge joining vertices $x$ and $y$ and $\Gamma(x)$ is the neighbourhood of $x$.
\end{definition}
The weighted distribution is straightforwardly obtained by using the
weighted degree in Definition~\ref{def:ndd}.

\begin{figure}
   \centering
   \includegraphics[width=\textwidth]{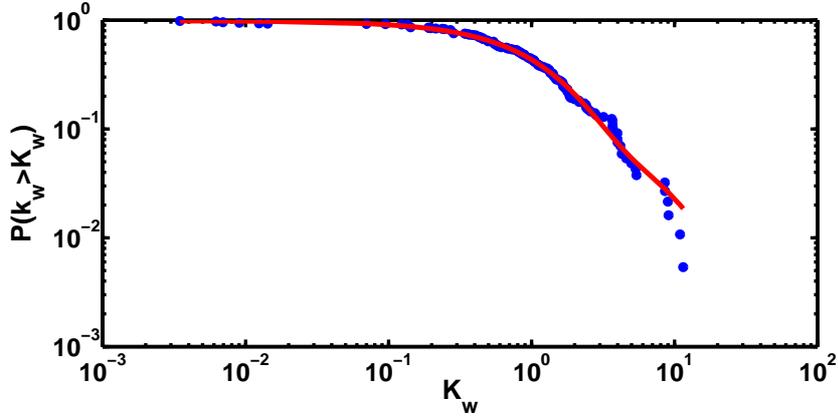}
   \caption{Weighted Node Degree Cumulative Probability Distribution for Low Voltage sample \#5 (logarithmic scale). Circles represent sample data, while continuous line represents a sum of exponential decays $y=0.8975e^{-0.9289x}+0.0904e^{-0.1379x}$.}
\label{fig:wnddLV5}
\end{figure}

\begin{figure}
   \centering
   \includegraphics[width=\textwidth]{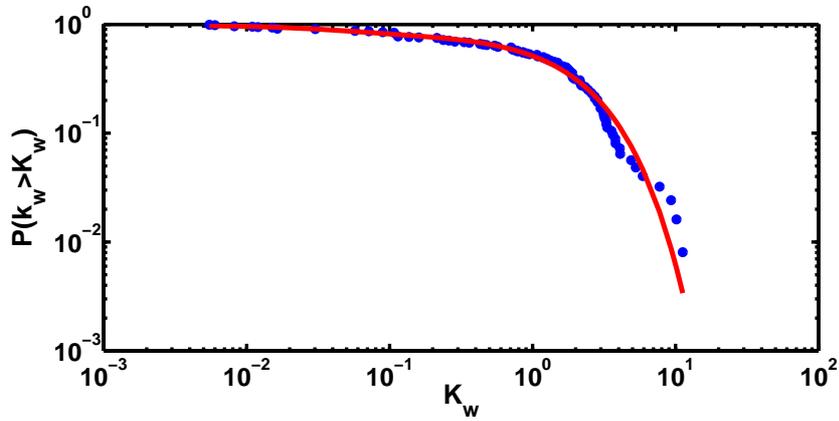}
   \caption{Weighted Node Degree Cumulative Probability Distribution for Low Voltage sample \#10 (logarithmic scale). Circles represent sample data, while continuous line represents a sum of exponential decays $y=0.1538e^{-21.47x}+0.8378e^{-0.4909x}$.}
\label{fig:wnddLV10}
\end{figure}

\begin{figure}
   \centering
   \includegraphics[width=\textwidth]{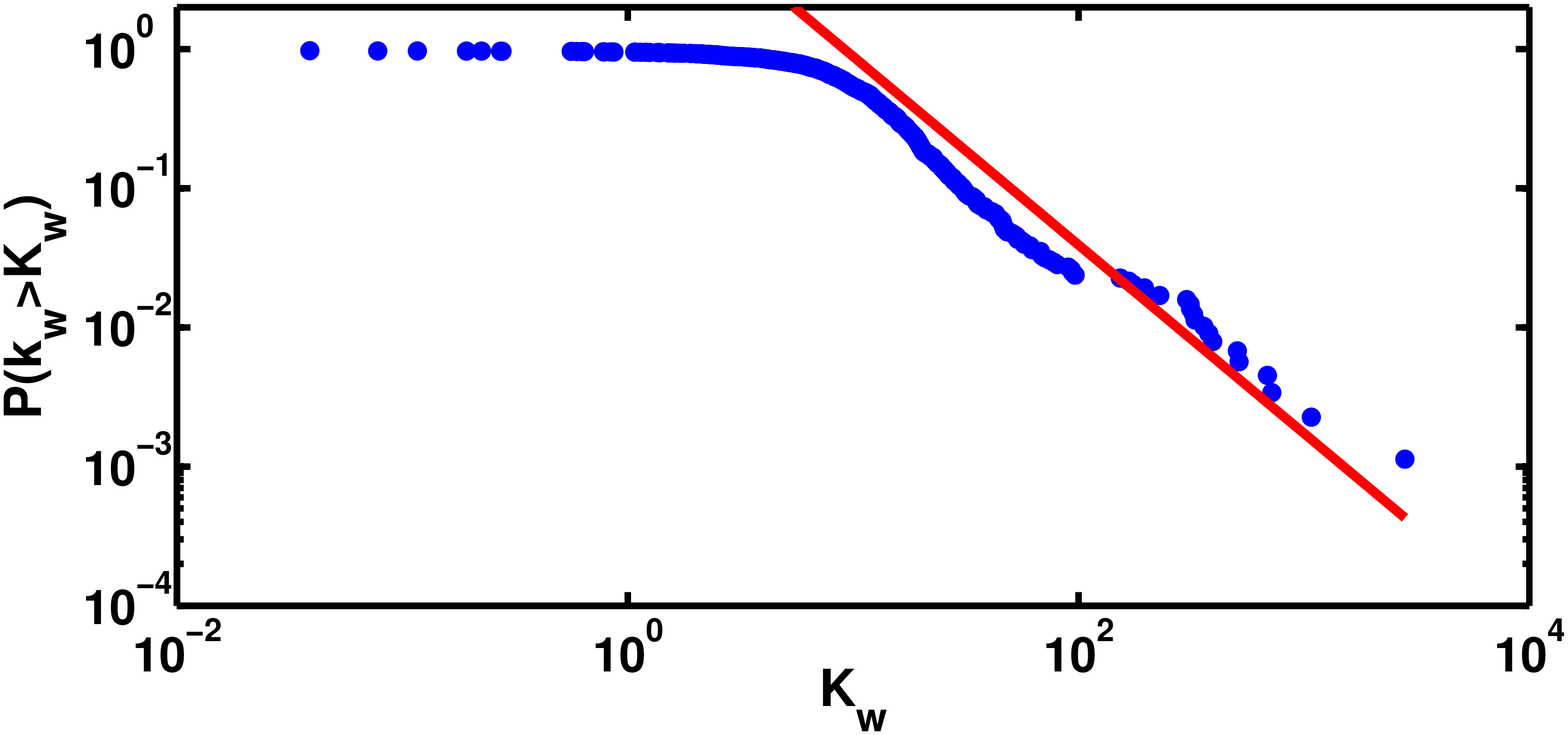}
   \caption{Weighted Node Degree Cumulative Probability Distribution for \MV sample \#2 (logarithmic scale). Circles represent sample data, while straight line represents a power-law with $\gamma=1.354$.}
\label{fig:wnddMV2}
\end{figure}

\begin{figure}
   \centering
   \includegraphics[width=\textwidth]{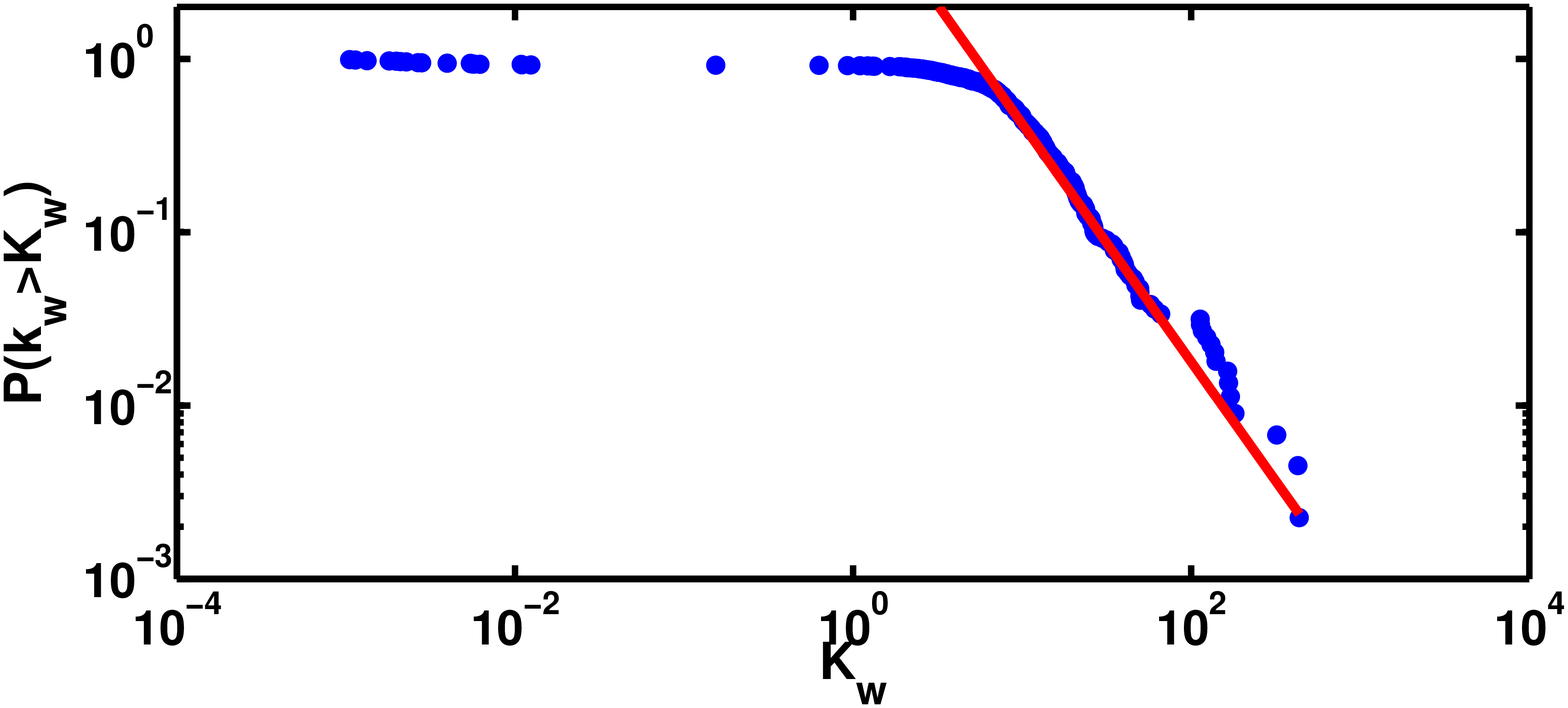}
   \caption{Weighted Node Degree Cumulative Probability Distribution for \MV sample \#3 (logarithmic scale). Circles represent sample data, while straight line represents a power-law with $\gamma=1.374$.}
\label{fig:wnddMV3}
\end{figure}

For the most significant sample of the Low Voltage, as shown in
Figures~\ref{fig:wnddLV5} and~\ref{fig:wnddLV10}, the shape of the
distribution is close to an exponential one with a quite fast
decay. The situation looks different in \MV samples. The very first
part of the distribution is well fitted by an exponential
shape, while the central part of the distribution, and
especially the tail, fit best a power-law like
shape as visible in Figures~\ref{fig:wnddMV2} and~\ref{fig:wnddMV3}. An
explanation of such behaviour between the the most numerous samples of
the two ends of the Grid is due to the order and size of the \MV
samples which are from two to four times bigger than the \LV samples, thus having a higher likelihood of far different values in weighted node degree.

\subsection{Betweenness}

\begin{figure}
   \centering
   \includegraphics[width=\textwidth]{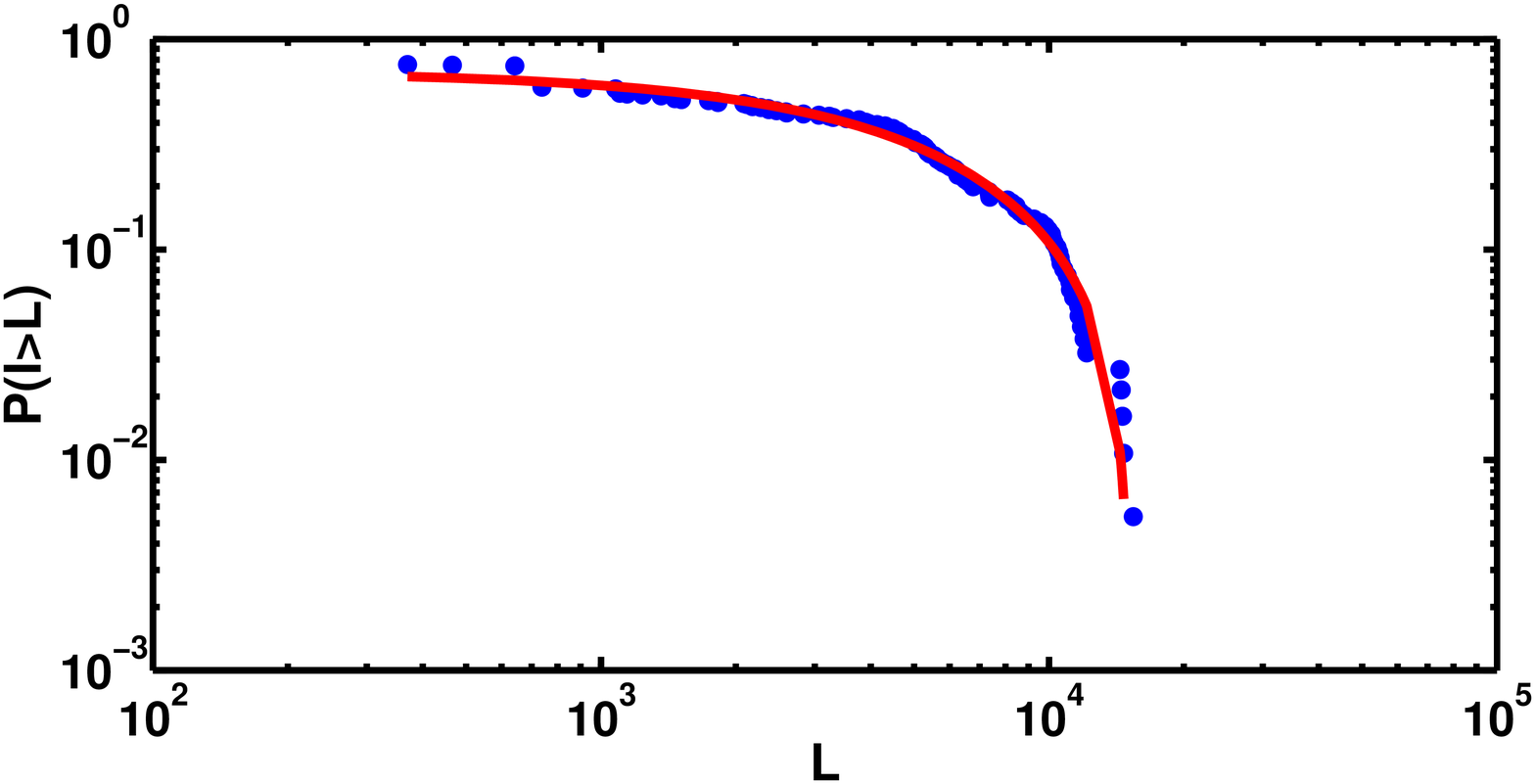}
   \caption{Betweenness Cumulative Probability Distribution for \LV sample \#5 considering the weighted graph (logarithmic scale).  Circles represent sample data, while continuous line represents a sum of exponential decays $y=-0.1051e^{3.381\cdot10^{-6}x}+0.8084e^{-1.317\cdot10^{-4}x}$.}
\label{fig:wbpdLV5}
\end{figure}

\begin{figure}
   \centering
   \includegraphics[width=\textwidth]{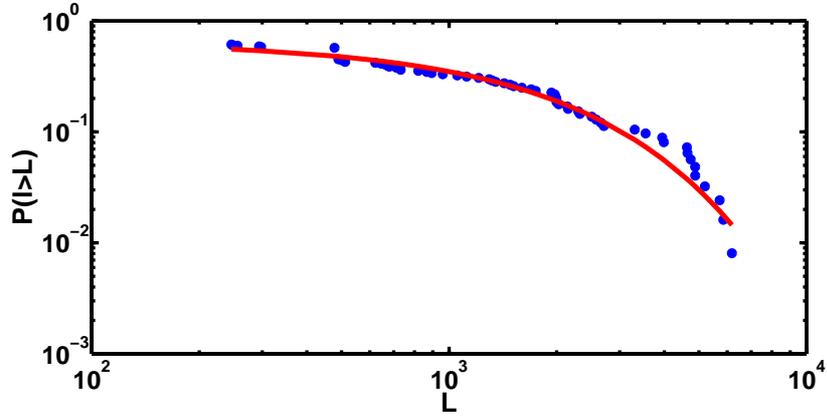}
   \caption{Betweenness Cumulative Probability Distribution for \LV sample \#10 considering the weighted graph (logarithmic scale). Circles represent sample data, while continuous line represents an exponential decay $y=0.6456e^{-6.139\cdot10^{-4}x}$. }
\label{fig:wbpdLV10}
\end{figure}

\begin{figure}
   \centering
   \includegraphics[width=\textwidth]{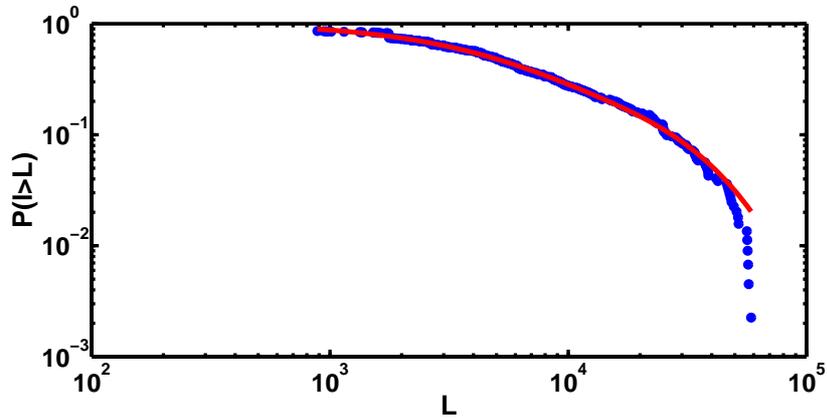}
   \caption{Betweenness Cumulative Probability Distribution for Medium Voltage sample \#3 considering the weighted graph (logarithmic scale). Circles represent sample data, while continuous line represents a sum of exponential decays $y=0.6582e^{-2.648\cdot10^{-4}x}+0.3939e^{-5.060\cdot10^{-5}x}$.}
\label{fig:wbpdMV3}
\end{figure}

Considering the weighted definition of path it is possible to compute betweenness in the weighted scenario and again betweenness of a node can be seen as a random variable thus obtaining the corresponding probability distribution. The shape of the distribution does not change much compared to the same unweighted
samples for the \LV network, as shown for samples \#5 and \#10 in
Figures~\ref{fig:wbpdLV5} and~\ref{fig:wbpdLV10}: the distribution is
best approximated by an exponential decay or by a sum of exponential
contributions. For \MV samples, the changes between unweighted and
weighted paths influence the betweenness probability distribution
whose shape in these conditions seems to be better approximated by an
exponential or sum of exponential components as shown in
Figure~\ref{fig:wbpdMV3}. This change in the distribution of the number of
shortest paths that traverse a node between the weighted and the
unweighted graph is clearly an indication that some property change
between the two analysis and it is worth to remember that the weighted
path analysis better approximate the actual routes current flows
follow.

\subsection{Node Importance}

The criticality or importance of nodes in the network is best studied
using weights. In fact, an edge with high capacity (i.e., weight) makes a node very important,
conversely many edges with little capacity (i.e., weight) make a node almost
irrelevant. The approach considered is similar to the one leading to
the eigenvector centrality computation performed in
Section~\ref{sec:unwRes}. However, to compute this metric, a weighted
form of the adjacency matrix is necessary (cf. Newman~\cite{newman04}).
\begin{definition}[Weighted adjacency matrix]
\label{def:WAdjMat}
The weighted adjacency matrix $A_w=A_w(G)=(a_{i,j})$ of a graph $G$ is the $n \times n$ matrix given by
\[
 a_{ij} =
  \begin{cases}
   w_{ij}      & \text{if } v_iv_j \in E \text{ and has weight } w_{ij}, \\
   0       & \text{otherwise.}
  \end{cases}
\]
\end{definition}
Computing the eigenvector corresponding to the principal eigenvalue, one
obtains a ranking among the various nodes of the network. The results for eigenvector centrality for two \MV samples are shown in Tables~\ref{tab:eigvcentralW5LV} and~\ref{tab:eigvcentralW10LV} while for two \MV samples are shown in Tables~\ref{tab:eigvcentralW2MV} and~\ref{tab:eigvcentralW3MV}. The first column of each table represents the ranking position for the first ten node whose identifier is given in the second column. It is
interesting to note that with this ranking the first ten most
important nodes of the graph, and to a certain extent critical
substations for the Power Grid, are in general different than those of the unweighted analysis. This aspect
reinforces a consistent difference in node properties between the two
type of analysis performed. 

\begin{table}[h!b!p!]
\begin{center}
\begin{footnotesize}
    \begin{tabular}{ | p{3cm} | p{3cm}|}
    \hline
    Eigenvector Centrality Ranking \# & Node ID \\ \hline
1&192\\ \hline
2&191\\ \hline
3&6\\ \hline
4&24\\ \hline
5&137\\ \hline
6&130\\ \hline
7&135\\ \hline
8&129\\ \hline
9&178\\ \hline
10&transformer 3\\ \hline
    \end{tabular}
    \end{footnotesize}
\end{center}
\caption{Eigenvector centrality ranking for \LV sample \#5 considering the corresponding weighted graph.}
\label{tab:eigvcentralW5LV}
\end{table}

\begin{table}[h!b!p!]
\begin{center}
\begin{footnotesize}
    \begin{tabular}{ | p{3cm} | p{3cm}|}
    \hline
    Eigenvector Centrality Ranking \# & Node ID \\ \hline
1&3\\ \hline
2&44\\ \hline
3&108\\ \hline
4&5\\ \hline
5&107\\ \hline
6&55\\ \hline
7&110\\ \hline
8&109\\ \hline
9&102\\ \hline
10&39\\ \hline
    \end{tabular}
    \end{footnotesize}
\end{center}
\caption{Eigenvector centrality ranking for \LV sample \#10 considering the corresponding weighted graph.}
\label{tab:eigvcentralW10LV}
\end{table}

\begin{table}[h!b!p!]
\begin{center}
\begin{footnotesize}
    \begin{tabular}{ | p{3cm} | p{3cm}|}
    \hline
    Eigenvector Centrality Ranking \# & Node ID\\ \hline
1&577\\ \hline
2&580\\ \hline
3&575\\ \hline
4&822\\ \hline
5&702\\ \hline
6&706\\ \hline
7&546\\ \hline
8&578\\ \hline
9&170\\ \hline
10&90\\ \hline
    \end{tabular}
    \end{footnotesize}
\end{center}

\caption{Eigenvector centrality ranking for \MV sample \#2 considering the corresponding weighted graph.}
\label{tab:eigvcentralW2MV}
\end{table}

\begin{table}[h!b!p!]
\begin{center}
\begin{footnotesize}
    \begin{tabular}{ | p{3cm} | p{3cm}|}
    \hline
    Eigenvector Centrality Ranking \# & Node ID\\ \hline
1&324	\\ \hline
2&351	\\ \hline
3&263	\\ \hline
4&350	\\ \hline
5&299	\\ \hline
6&410	\\ \hline
7&393	\\ \hline
8&48	\\ \hline
9&124	\\ \hline
10&80	\\ \hline
    \end{tabular}
    \end{footnotesize}
\end{center}
\caption{Eigenvector centrality ranking for \MV sample \#3 considering the corresponding weighted graph.}
\label{tab:eigvcentralW3MV}
\end{table}

\subsection{Fault tolerance}

\begin{figure}
   \centering
   \includegraphics[width=\textwidth]{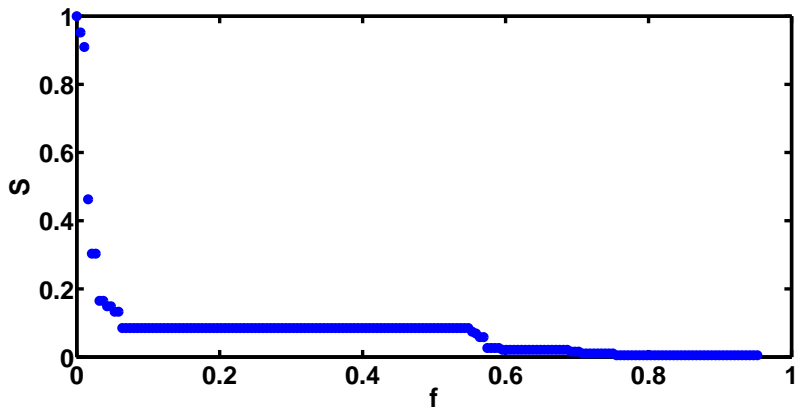}
   \caption{Resilience for weighted node degree-based removal for \LV sample \#5. The horizontal axis represents the fraction $f$ of the nodes removed from the original sample; the vertical axis represents the size of the largest connected component $S$ relative to the initial size of the graph.}
\label{fig:wremLV5}
\end{figure}

\begin{figure}
   \centering
   \includegraphics[width=\textwidth]{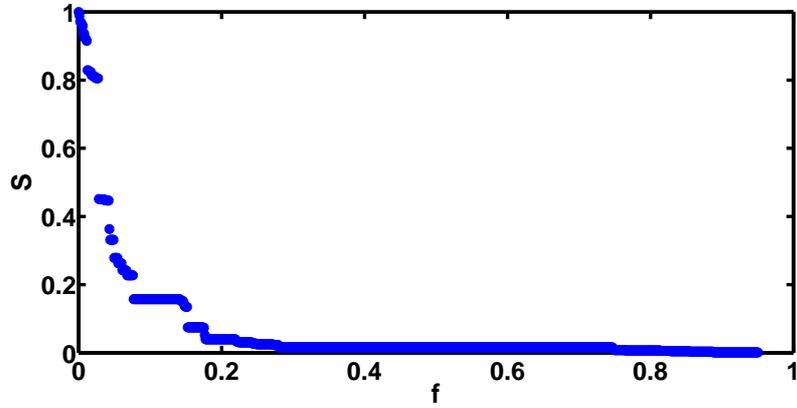}
   \caption{Resilience for weighted node degree-based removal for \MV sample \#2. The horizontal axis represents the fraction $f$ of the nodes removed from the original sample; the vertical axis represents the size of the largest connected component $S$ relative to the initial size of the graph.}
\label{fig:wremMV2}
\end{figure}

\begin{figure}
   \centering
   \includegraphics[width=\textwidth]{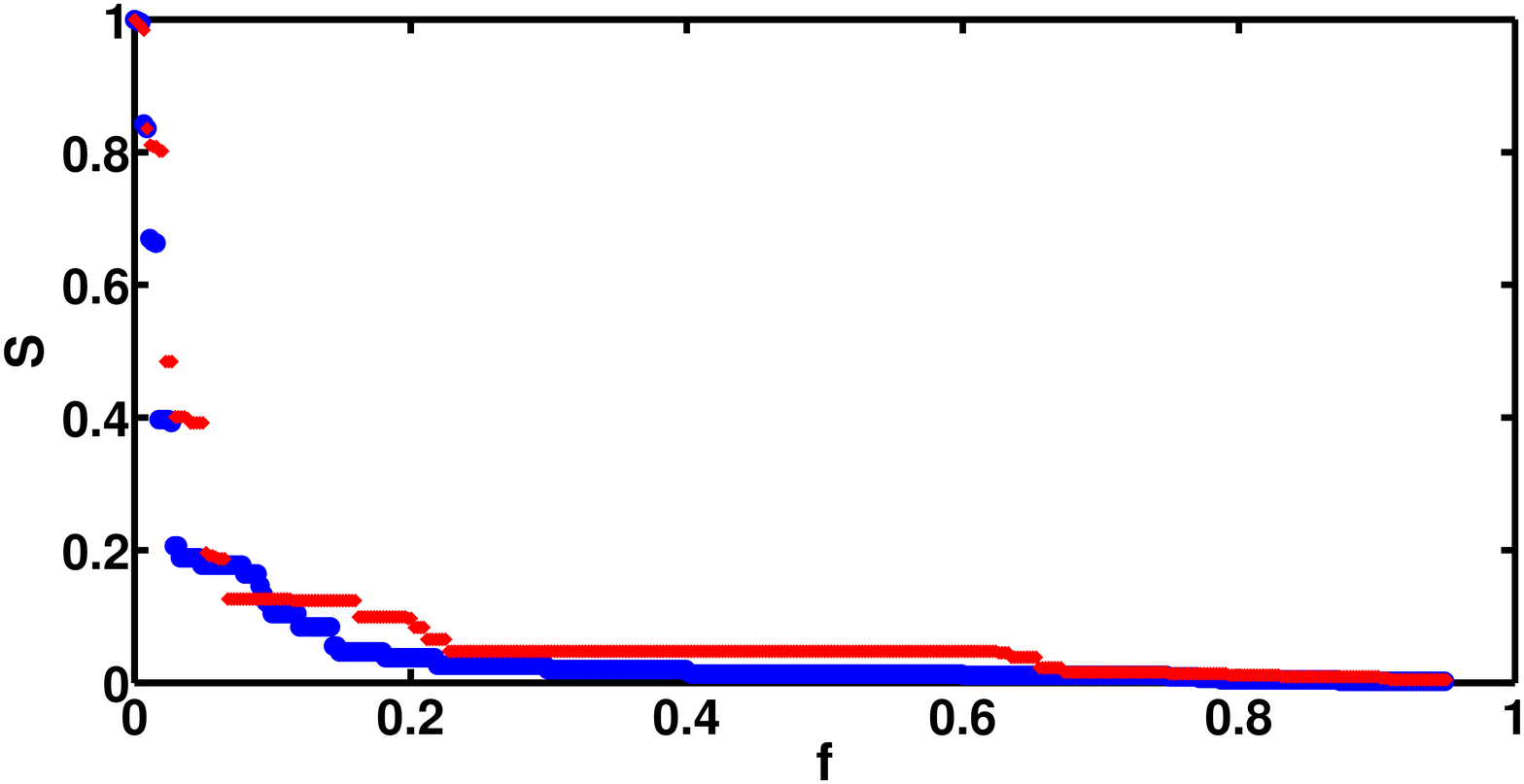}
   \caption{Resilience for node degree-based removal for \MV sample \#3. The horizontal axis represents the fraction $f$ of the nodes removed from the original sample; the vertical axis represents the size of the largest connected component $S$ relative to the initial size of the graph. Red diamonds represent the weighted node degree-based removal, while blue circles represent \textit{traditional} node degree-based removal.}
\label{fig:compremMV3}
\end{figure}

\begin{figure}
   \centering
   \includegraphics[width=\textwidth]{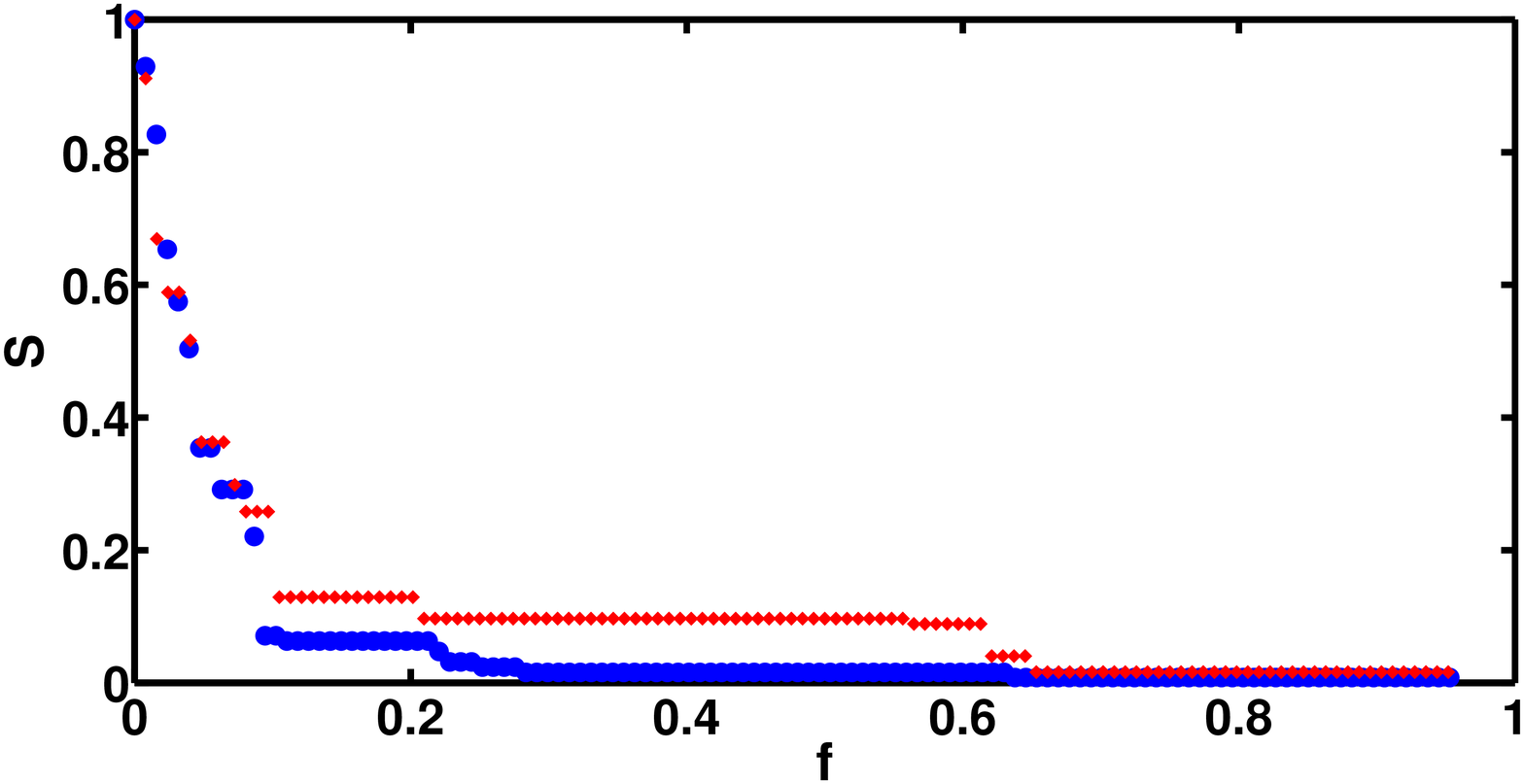}
   \caption{Resilience for node degree-based removal for \LV sample \#10. The horizontal axis represents the fraction $f$ of the nodes removed from the original sample; the vertical axis represents the size of the largest connected component $S$ relative to the initial size of the graph. Red diamonds represent the weighted node degree-based removal, while blue circles represent \textit{traditional} node degree-based removal.}
\label{fig:compremLV10}
\end{figure}

Fault tolerance can be evaluated based on the removal of nodes
following strategies similar to the unweighted case. Since the
random removal yields exactly the same result for the weighted and
unweigted case, here we focus on the
node degree-based removal policy which considers the weighted node
degree definition. The disruption behaviour of the network samples is very
similar to the unweighted situation: the network suffers deeply these targeted attacks; a
very small percentage of removed nodes causes an important loss
in the size of the biggest component left in the network. The
comparison between Figures~\ref{fig:wremLV5} and~\ref{fig:remNDLV5} for
the \LV samples, and Figures~\ref{fig:wremMV2} and~\ref{fig:remNDMV2}
for the \MV samples provides a general correlation between high
degree nodes in the unweighted graph and high degree nodes in the
weighted one. If one takes a closer look at the disruption charts for
the same samples some small differences can anyway be noticed. The
nodes with the highest weighted degree cause a bigger
damage to the network when removed in the very first iteration than nodes
with higher degree in unweighted networks, this behaviour is shown in 
Figures~\ref{fig:compremMV3}  and~\ref{fig:compremLV10}. The situation then
changes in the later stages of the removal process when a bigger
disruption is caused by nodes with higher node degree in
\textit{traditional sense}.

\section{Unweighted vs. Weighted comparison study}\label{sec:unweightedweightedcomparison}

The weighted study of the \PG presented in the previous section
has already highlighted the more precise information available with
such study. Next we consider more in detail the comparison between the
unweighted and weighted study for the most indicative measures.

\begin{figure}[htbp]
 \centering
 \resizebox{\textwidth}{!}{ \includegraphics{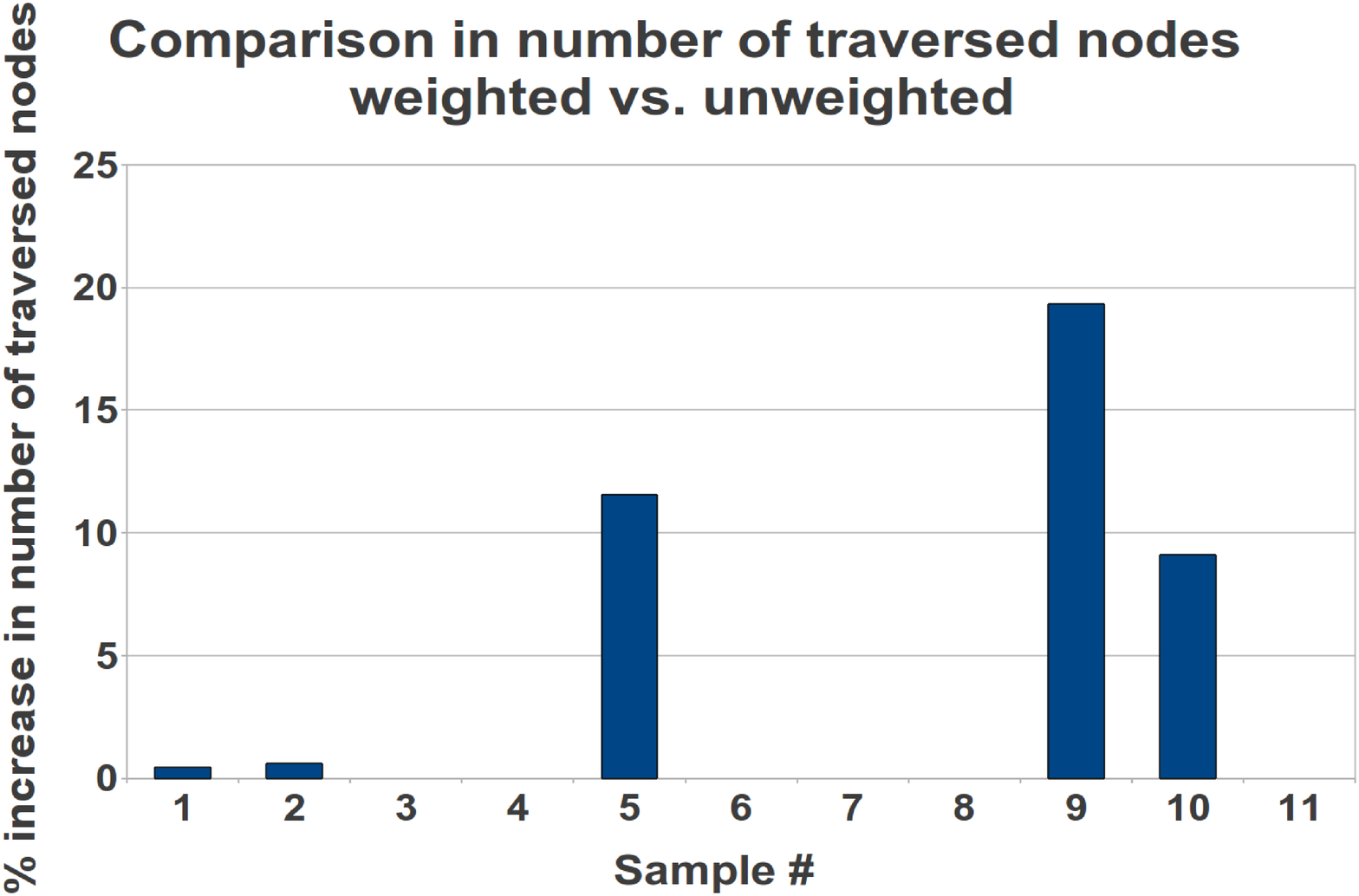}}
 \caption{Percentage increase in number of node traversed between weighted and un-weighted \LV samples.}
 \label{fig:increaseLV}
\end{figure}

\begin{figure}[htbp]
 \centering
 \resizebox{\textwidth}{!}{ \includegraphics{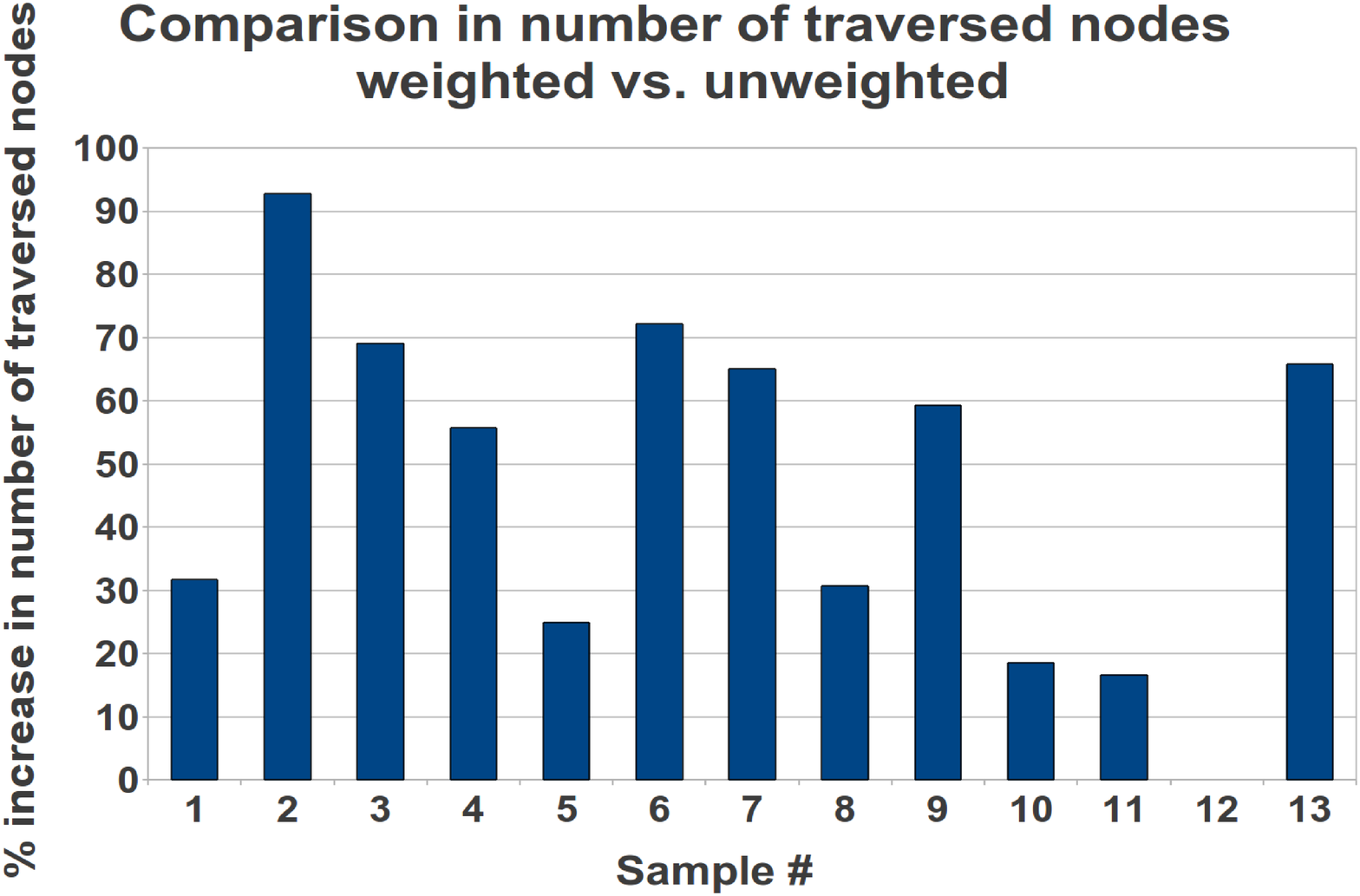}}
 \caption{Percentage increase in number of node traversed between weighted and un-weighted \MV samples.}
 \label{fig:increaseMV}
\end{figure}

Considering a minimal path, one may wonder if
introducing weights changes the number of traversed nodes by such
minimal paths on average. Even if the WCPL and the NWCPL values are close, there might be differences in
the number of nodes and (therefore edges) traversed when following a
minimal path. This is particularly interesting from the practical
point of view, as it indicates the number of transformers and
distribution substations traversed in the Power Grid. These points are
critical in terms of additional losses that are associated with
substations and transformers, and in turn in the number of potential
points of failure that a path traverses. Figures~\ref{fig:increaseLV}
and~\ref{fig:increaseMV} show the results for the \LV and Medium
Voltage, respectively. Each bar represents the average percentage
increase in the number of nodes traversed along the shortest path
between any two nodes for the unweighted and the weighted
situation. It is interesting to note that for several samples of
the \LV there is no difference in the number of traversed nodes, thus
reinforcing the idea of a highly hierarchical tree-like structure whose paths are fixed by the built-in topology of the Grid independently of the associated edge weights. The
situation though is quite different for the Medium Voltage. In fact there
is an increment of traversed nodes between the weighted and unweighted
models (especially for the meaningful samples) on average of about
50\%. This is a clear indication of a meshed network for which there
are less imposed paths and in which the weights have an important role. It is important to notice that for the
biggest sample (almost 900 nodes) the number of visited nodes when
following a path between any two nodes increases by more than 80\%
comparing the unweighted and the weighted situation.

Considering the node degree distribution it seems that the weighted
analysis tends to reduce the contribution of the tail components of the
distribution, thus being more compact especially for \LV samples.
This is due to the small variance of the weighted node degree that especially these low-end samples of the Grid show.
Samples that in the
unweighted analysis show a power-law distribution, when considered
weighted tend to assume an exponential form or a sum of exponential
contributions. The biggest samples analysed in the \MV (samples \#2
and \#3) tend to even have a faster decay than the unweighted
situation. The same consideration applies
to the betweenness probability distribution which for the \LV samples
still shows a best approximation by an exponential decay. For the samples belonging to the \MV the same
tendency appears: a more compact distribution of the number of paths
traversing nodes.

Considering a representation of the relationships between nodes and
their weights,
the ranking of importance obtained with the eigenvector centrality computation between nodes changes
substantially. This is understandable since the weight deeply
influences the properties, and therefore importance, of a node. This
dissimilarity in node characteristics is also shown by the different
behavior the removal of the highest degree nodes brings to the network
connectivity. Although the general behavior functioning similar, the first nodes removed with the highest weighted
degree tend to have a more damaging impact than the corresponding
policy removal in an unweighted graph, this behavior then tends to
reverse while removing more and more nodes: the most damaging results
in term of network connectivity is then caused by nodes with highest
degree in the unweighted network.
The comparison between weighted and unweighted node degree-based
removal is shown in Figure~\ref{fig:compremLV10} for a \LV sample and in Figure~\ref{fig:compremMV3} for a \MV
sample.

\section{Topological influence on Energy Exchanges}\label{sec:economic}

Traditionally, energy has been `pushed' hierarchically from the large
scale production facilities to the end users. Famous is the quote of
Samuel Insull (XIX century):
\begin{quote}
{\em 
``every home, every
factory and every transportation line will obtain its energy from one
common source, for the simple reason that that will be the cheapest
way to produce and distribute it.'' 
}
\end{quote}
 Clearly, things are rapidly changing on today's \PG and new models are
 emerging where delocalized production is the norm, rather than the
 exception. The trend will call for an infrastructure that supports energy
 trading among any actor connected to the Power Grid. 

The Complex Network Analysis that we provided so far gives a
statistical aggregated view of the current infrastructure for the Low
and Medium Voltage Grids. The natural next question that arises concerns the
usability of such infrastructure for the delocalized energy
exchange. To answer it, we propose to tie statistical
properties of the \PG with a trading cost. The cost represents a
balance frontier below which the actors are motivated to trade and
above which they are not. It is important to remark that we do not claim of having
identified ``the'' cost function, but rather we propose that Complex
Network Analysis measures are deeply connected to the success of a
decentralized energy market.

\subsection{Relating topology to price}

In general, establishing energy pricing is not a simple task since several aspects
influence the overall price at which electricity is sold. There are
aspects connected to the supply side such as fuel prices, policy regulations, load losses and bidding
strategies; on the other hand, there are elements connected to the
demand side such as human behaviors, natural phenomena that 
influence habits and thus consumption. Recent proposals and methods for price allocation 
include {\em nodal pricing}~\cite{Leuthold2008284},
which is particularly indicated for distributed generation solutions because of the
price benefits it brings to the
customer~\cite{Sotkiewicz2006}. It is also important to notice that
the savings deriving from distribution
losses can be extremely important~\cite{Chiradeja2005,Khoa2006,Le2007}. 
 A set of factors is most
definitely tied to infrastructural properties of the distribution
network, as illustrated for instance in the economic studies of Harris
and Munasinghe~\cite{Harris06,Munasinghe84},
most notably:
\begin{itemize}
\item losses both in line and at transformer stations,
\item security and capacity factors,
\item line redundancy, and
\item power transfer limits.
\end{itemize}
The listed technical parameters are naturally associated with a
topological parameter, namely:
\begin{itemize}
\item Line losses are related to and thus expressed as a function of the weighted
  characteristic path length $L_{line_N} = f(WCPL_N)$

\item Substation losses are expressed as a function of the (average)
   number of nodes visited while travelling from source node to
  destination node along the weighted 
  shortest path $L_{substation_{ij}} = f(|WSP_{ij}|)$. The significant dimension is $L_{substation_N} =
  f(\overline{|WSP_N|})$ where $\overline{|WSP_N|}$ is the average
  number of nodes traversed in a shortest path in the network $N$.

\item Robustness from a topological perspective is expressed as the fraction of the maximal connected
  component compared to its original size once a certain
  fraction of nodes is removed. $Rob_N = f(N,p)$ where $N$ is the network under
  evaluation and $p$ is the removal policy adopted.

\item Line redundancy is simply mapped to a topological metric
  that counts the number of paths (without cycles) that are available
  between any two nodes  and the cost associated to this redundancy $Red_{ij} = f(|P_{ij}|, w_{ij})$ where $P_{ij}$ is the set
  of paths between nodes $i$ and $j$ and $w_{ij}$ is the weight
  of the worst case redundant path. A global metric for network $N$ is
 $Red_{N} = f(\overline{|P_{ij}|}, \overline{w_{ij}}), \forall i,j \in N$

\item Network capacity may be considered as a function of the weighted
 characteristic path length where the weight is the maximal supported operating
 current of the cable $Cap_{N} = f(WCPL_N)$
\end{itemize}
\begin{figure}  
 
   \centering
   \includegraphics[width=\textwidth]{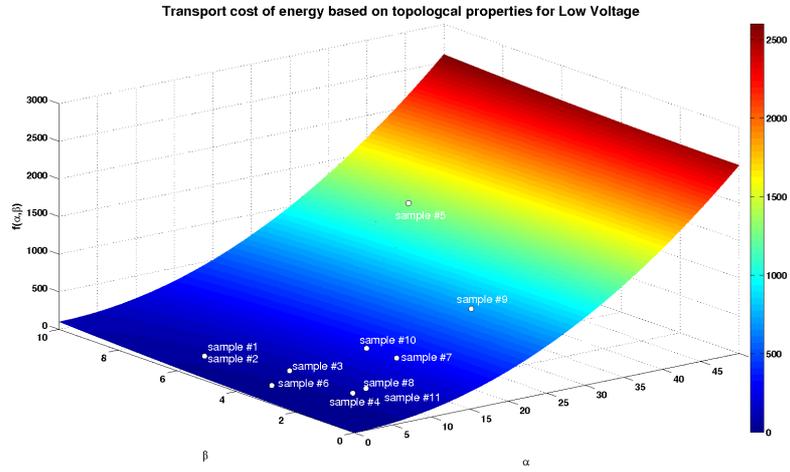}
   \caption{Transport cost of energy based on the topological properties for Low Voltage samples.}
\label{fig:priceLV}
\end{figure}
\begin{figure}
 
   \centering
   \includegraphics[width=\textwidth]{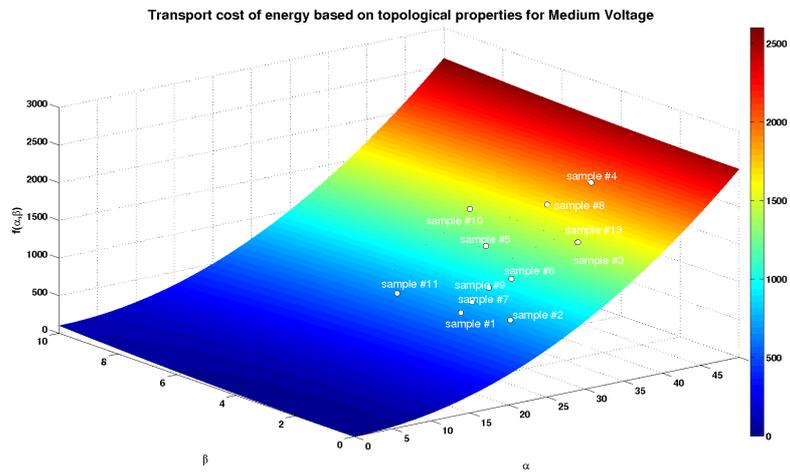}
   \caption{Transport cost of energy based on the topological properties for Medium Voltage samples.}
\label{fig:priceMV}
\end{figure}

These topological ingredients provide two sorts of measures, the first
one $\alpha$ gives an average of the dissipation in the transmission between
two nodes 
\begin{align} 
\alpha & = f(L_{line_N},L_{substation_N})\label{eq:alpha};
 \end{align} the second one $\beta$ is a measure of reliability/redundancy in
the paths among any two nodes
\begin{align}
 \beta & = f( Rob_{N},Red_{N},Cap_{N})\label{eq:beta}.
 \end{align}
We argue that these two factors
influence the inclination of prosumers (energy consumers/producers) to
trade energy on the Power Grid.
In fact a high value of the $\alpha$ parameter represents a high level
of losses experienced for transporting energy in the network, either in distribution lines or substations.
Additionally, the reliability and ability to bring sufficient energy
to the end users plays an important role. In fact, if proper levels of
robustness of the network or resilience to failures are not the norm,
the prosumer inclination to sell energy as well as the end user to buy it
will be limited.
Furthermore, if the availability of redundant paths for electricity
routing in case of partial disruption of the network are insufficient
leads to a high value of
$\beta$, and in turn a disincentive for trading. To better understand
the constituents of $\alpha$ and $\beta$, we consider next a possible
instantiation of these parameters using the data of the Dutch Power Grid.

\subsection{An example}

To give an
impression of how the parameters can be used to assess the success of
the energy market, we provide an example next. We stress that this is
simply an example, and it does not have the ambition to provide the
successful parameter for the delocalization of the energy distribution
market.

\begin{itemize}
\item Losses on the transmission/distribution line can be expressed by the quotient of the weighted
  characteristic path length and the average weight of a line (a weighted edge in the graph):
\begin{center}$L_{line_N} = \frac{WCPL_N}{\overline{w}}$\end{center}

\item Losses at substation level are expressed as the number of nodes (on average) that are traversed when computing the weighted shortest path between all the nodes in the network:
\begin{center}$L_{substation_N} = \overline{Nodes_{WCPL_N}}$\end{center}

\item Robustness is evaluated with random removal strategy and the weighted node degree-based removal by computing the average of the order of maximal connected component between the two situations when the 20\% of the nodes of the original graph are removed. It can be written as:
\begin{center}$Rob_N = \frac{|MCC_{Random20\%}|+|MCC_{NodeDegree20\%}|}{2}$\end{center}

\item Redundancy is evaluated by covering a random sample of the nodes
  in the network (40\% of the nodes whose half represents source nodes
  and the other half represents destination nodes) and computing for
  each source and destination pair the first ten shortest paths of
  increasing length. If there are less than ten paths available, the
  worst case path between the two nodes is considered. To have a
  measure of how these resilient paths have an increment in
  transportation cost, a normalization with the weighted
  characteristic path length is performed. We formalized it as:
\begin{center}$Red_{N} = \frac{\sum_{i \in Sources, j \in Sinks}{SP_{w_{ij}}}}{WCPL}$\end{center}

\item Network capacity is considered as the value of the weighted characteristic path length, whose weights are the maximal operating current supported, normalized by the average weight of the edges in the network (average current supported by a line). That is:
\begin{center}$Cap_{N} = \frac{WCPL_{current N}}{\overline{w_{current}}}$\end{center}
\end{itemize}
With these instantiations, equations (\ref{eq:alpha}) and (\ref{eq:beta}) become:
\begin{align}
\alpha & = f(L_{line_N},L_{substation_N}) = L_{line_N}+L_{substation_N}\label{eq:alphaEx}\\
\beta & = f(Red_{N},Rob_{N},Cap_{N}) = \frac{Red_N}{Rob_N \cdot ln(Cap_N)}\label{eq:betaEx}
\end{align}

The functions to compute $\alpha$ and $\beta$ are only few
of many available possible ones. The choice made
here is to have a simple mechanism to assess the potential
exchange costs of different networks.
Equation~(\ref{eq:alphaEx}) is a basic sum over the losses that are
experienced both at line and at substation level. 
Equation~(\ref{eq:betaEx}) takes into account the aspects of reliability and
tolerance of the network: the higher the $\beta$ the more prone
to failures and less reliable the network is. 
The cost of increasing the number of  paths to provide more redundancy is the dividend
in the fraction, while elements improving reliability act as
divisor.

\begin{figure*}
   \centering
   \includegraphics[width=1\textwidth]{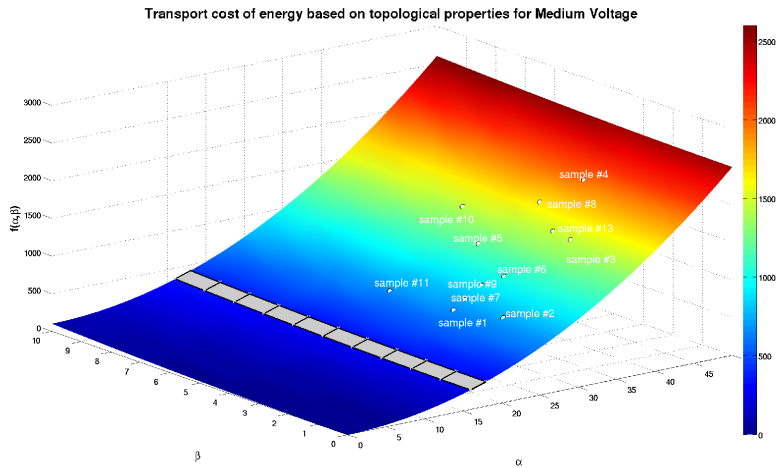}
   \caption{Transport cost of energy based on the topological properties for Medium Voltage with supposed economic convenience threshold (grey thick line).}
\label{fig:priceMVThreshold}
\end{figure*}

With these quantities, one can form an impression of what the influence of
the cost of transportation is for the decentralized energy
exchange. If the cost is too high because of an infrastructure with
high chances of failure (high $\beta$) and high resistance (high
$\alpha$), then the decentralized market will not be incentivized. On the
other hand, for low $\alpha,\beta$, it will be economically attractive to
have a decentralized energy market. In Figures~\ref{fig:priceLV} and~\ref{fig:priceMV}, we show
a combination of $\alpha,\beta$ obtained with the functions described above and with an hypothesis of a quadratic increase of energy price with the increment in $\alpha$ and $\beta$. We report the position of the analysed samples as white circles in Figures~\ref{fig:priceLV} and~\ref{fig:priceMV} respectively for \LV and \MV samples. By performing an economic study, which we stress is beyond the scope of
the present treatment, one then can identify what the threshold is for the
feasibility of a decentralized market (grey area in Figure~\ref{fig:priceMVThreshold}) and
then conclude what topological modifications are necessary to the \MV
and \LV infrastructure in order to allow the energy exchange.

\section{Related Work}\label{sec:relwks}

Complex Network Analysis studies are becoming more and more popular
given the amount of natural and human based complex systems. The \PG
is clearly amenable to such study and a number of these have been
performed on the \HV Grid around the globe. These have been an inspiration for the
present work which though is novel in its analysis of the \MV and Low Voltage,
in its use of weights, and its motivation stemming from energy
negotiation by micro/medium prosumers. Next we survey the most notable
related work.

\begin{table*}[htb]

\begin{center}
\hspace*{-3.4cm}
\begin{footnotesize}
    \begin{tabular}{| c| c | c  | c | c | c | c |c| c|}
    \hline
    Work & Sample  & Sample & Network & Node Degree  & Betweenness  & Weighted/Unweighted & Resilience  & Small-world \\ 
    &Order & Type& Type& Distribution  &Distribution & Analysis &Analysis &  Investigation\\
     & & & &  Statistics   &Statistics&  & &\\\hline
    Albert \etal  & $\sim$14000 & Real & HV & \checkmark & \checkmark & Unweighted & \checkmark & \\ 
    in \cite{alb:str04}& & & && & & & \\ \hline
    Crucitti \etal & $\sim$300 & Real& HV & \checkmark & \checkmark & Weighted not based  & \checkmark & \\ 
    in \cite{Crucitti04} & & & & & & on physical properties & & \\\hline
    Chassin \etal  & $\sim$314000 & Real& HV & \checkmark &  & Unweighted & \checkmark & \\ 
    in \cite{Chassin05}&&&&&&&& \\ \hline
    Holmgren \etal & $\sim$4800 & Real& HV & \checkmark &  & Unweighted & \checkmark & \checkmark\\ 
     in \cite{holmgren06}&&&&&&&& \\ \hline
    Casals \etal  & $\sim$2800 & Real& HV & \checkmark &  & Unweighted  &  & \\ 
    in \cite{Rosas-Casals2009}&&&&&&&& \\ \hline
    Casals \etal  & $\sim$3000& Real & HV & \checkmark &  &  Unweighted & \checkmark & \checkmark \\ 
    in\cite{casals07}&&&&&&&& \\ \hline
    Sole \etal  & $\sim$3000 & Real& HV & \checkmark &  &  Unweighted & \checkmark & \\ 
    in\cite{Corominas-murtra2008}&&&&&&&& \\ \hline
    Crucitti \etal  & $\sim$130 & Real& HV &  &  &  Unweighted & \checkmark & \\ 
    in\cite{Crucitti2005}&&&&&&&& \\ \hline
    Rosato \etal  & $\sim$130& Real & HV & \checkmark &  &Unweighted  & \checkmark & \\ 
    in\cite{Rosato2007}&&&&&&&& \\ \hline
    Watts \etal  & $\sim$4900& Real & HV &  &  &Unweighted  &  & \checkmark\\ 
    in\cite{Watts03}&&&&&&&& \\ \hline
    Wang \etal & $\sim$8500& Synthetic & HV &\checkmark  &  &Unweighted   &  & \checkmark\\ 
    in\cite{Wang2010}&&and real&&&&and impedence analysis&& \\ \hline
    \textbf{Present Study} & \textbf{$\sim$4850}  & \textbf{Real}&\textbf{MV/LV} & \textbf{\checkmark} & \textbf{\checkmark} & \textbf{Both} & \textbf{\checkmark} & \textbf{\checkmark} \\ \hline
    
    \end{tabular}
    \end{footnotesize}
\end{center}

\caption{Comparison table between literature studies using Complex Network Analysis applied to \PG networks.\label{tab:compa}}
\end{table*}

Albert \textit{et al.}~\cite{alb:str04} study the reliability aspects of the United States Power Grid. 
In extreme summary this work is particularly relevant for the big sample it analyses representing the whole North American \HV Power Grid and focusing in illustrating the cascading effects evaluating a connectivity loss property the authors define. The effects affecting the \PG based on this metric show important differences between the various policies of node removal (random, node degree or betweenness based). Betweenness computation is remarkable as it is used to identify the important nodes in the network, however the article does not take into account any sort of weight to be associated to the power lines.

Crucitti \textit{et al.}~\cite{Crucitti04} analyse the Italian \HV \PG from a topological perspective.
This work is particularly relevant for the concept of efficiency that is used to understand the performance of the network. This metric is evaluated as a function of the tolerance of load both for edges and nodes. It is interesting that some sort of weights are used for this analysis: a capacity measure is associated with nodes while weight is associated with edges based on residual capacity of nodes. In this article too, some strategies of failure simulation are taken into account (random and betweenness-based removal). However, the size of the sample analysed is small compared to other works and the type of weight attributed to edges is not related to any physical quantity (e.g., lines resistance), but is only related to topological betweenness.

A different model is presented by Chassin \textit{et al.}~\cite{Chassin05} where the analysis focuses on the North American Power Grid. The authors start with the hypothesis that the Grid can be modelled as a Scale-free network. 
This work is extremely relevant for the large size of the sample analysed (more than 300000 nodes) and the use of reliability measures that are typical of power engineering (e.g., loss of load probability) to quantify the failure characteristics of the network from a topological point of view. The similarity of the results obtained by considering reliability with authors' topological measures and other non topological studies for electrical Grids is interesting. However, a study of betweenness is unfortunately missing. Given the  size of the sample (although it is not explicitly stated if \MV components are considered or not) it would have useful to compute the betweenness of the nodes in order to understand if and how the betweenness behaves in this sort of big sample.

Holmgren~\cite{holmgren06} analyses the Nordic \PG involving the \HV Grids of Sweden, Finland, Norway, and a great part of Denmark comparing these with the U.S. Power Grid.
A resilience analysis is performed and  with the inclusion in this work of some fictitious scenarios of failure of the Grid and the possible adoptable solutions together with their resulting benefits. A computation of the betweenness property of the graph might have been useful to understand the differences between the different samples; also a weighted graph study might have pointed out even more interesting aspects between the various networks, but these are missing.

Casals \textit{et al.}~~\cite{Rosas-Casals2009} analyse the whole \HV European \PG and try to extract non-topological reliability measures investigating the topological properties of the network. The \PG analysed is the \HV end composed of almost 2800 nodes that span across the European continent. The assumption is that node degree distribution follows an exponential decay for every single network composing the European network, each one having a characteristic parameter that is related to the \emph{robustness} of the specific Grid. 
Although this study is based on relevant samples, but half of the considered Grids are small both in size and order (below 100 nodes). The most interesting aspect is the use of new indicators to assess network reliability, but, as the same authors explicitly state, these metrics need more test and a deeper study. As remarked for other works, there is no mention about using weights to characterize the edges in the networks.

Casals \textit{et al.}~\cite{casals07} consider the \HV Grids of many European countries analyzing them together and as separate entities.
This work has an overall relevant sample although no information are given for each single Grid which might have smaller significance when analysed alone. The most interesting aspects are the evaluation of Small-world properties for the networks composing the European Grid and the resilience test with both random and node degree-based removal strategies. There is no mention about using weights to characterize the edges in the network and no betweenness computation to find out critical nodes in term of paths covered.

Sole \textit{et al.}~\cite{Corominas-murtra2008} go further in exploring the same \PG data analysed in~\cite{casals07}, in particular they focus on targeted attacks. The model is based on the assumption, also verified by empirical data, that there is no correlation between nodes having a certain degree to be connected to each other. 
This work has an overall relevant sample and it focuses on simulating different failure events (random or targeted), establishing an interesting correlation between topological and non-topological reliability studies. The major point of improvement is related to the small size of half of the samples used (below 100 in order) and the possibility of introducing weights for edges related to some of the physical properties. In addition,  an evaluation of the betweenness is missing in order to understand if other nodes appear to be critical and so to be targeted using this different removal metric.

Crucitti \textit{et al.}~\cite{Crucitti2005} analyse the \HV \PG of Italy, 
France 
and Spain 
to detect the most critical lines and propose solutions to address their vulnerabilities. 
This work has some very valuable aspects such as the comparison of the Grids of three different countries (i.e., Italy, France and Spain), the identification of the most vulnerable edges, the damage provoked by an attack and possible improvements based on the efficiency metric. The sample used is considerably small; in addition there is no use of weights to characterize the edges. Thus it is not possible to discover which is the weight of the most critical edges identified and if there is a correlation with the unweighted analysis.

Rosato \textit{et al.}~\cite{Rosato2007} analyse the same network samples studied in~\cite{Crucitti2005} to investigate the main topological properties of these Grids (i.e., Italian, French and Spanish \HV Grids). 
The contributions of this work include the comparison of the Grid of three different countries, the identification of the most vulnerable edges, and the damage of an attack and achievable improvements based on adding edges. It also studies the node degree distribution and the shortest path length distribution for these samples. It is interesting to note how the authors clearly show the correlation between country geography and topological measures. The sample used is notably small; in addition there is no use of weights to characterize the edges.

Watts~\cite{Watts03} dedicates a subsection of his book to exploring the properties of the Western States Power Grid. 
The study gives motivations to the Small-world modeling. The analysis focuses on specific metrics such as network contraction parameters and the comparison between different models (i.e., relational and dimensional models). Therefore being Small-world the focus of the analysis, other typical Complex Network Analysis measuring are not performed (e.g., node degree distribution, betweenness distribution). 


Wang \textit{et al.}~\cite{Wang2010} study the \PG to understand the kind of communication system needed to support the decentralized control required by the Smart Grid. The analysis is based both on real \PG samples 
and synthetic reference models belonging to the IEEE literature.
This work has some very valuable aspects such as the investigation of a significant sample, the individuation of a new model for the node degree probability distribution, and the investigation of the physical impedance distribution of the Grid samples. All these factors bring to the development of a new model to characterize the Power Grid. An aspect that might have been analytically evaluated is the path length in the various samples analysed and the betweenness computation to characterize even those distribution analytically. The use of electrical properties is extremely interesting, however the analysis performed is dissociated to the physical graph properties therefore not considering a weighted graph structure.

There are also some brief studies related to the Power Grid that appear as examples in more general discussions about Complex Networks. In particular, Amaral \etal in~\cite{Amaral2000} show a study of the Southern California \PG and the model following an exponential decay for node degree distribution. Watts and Strogatz in~\cite{Watts98} show the Small-world phenomenon applied to the Western States Power Grid while Newman, within a more general work~\cite{Newman2003}, shows the exponential node degree distribution for the same Grid, while Barabasi \textit{et al.}~\cite{Barabasi1999a} model the \PG as a Scale-free network characterized by a power-law node degree distribution.

As shown in Table \ref{tab:compa} the analysis performed in the
present study has a sample significant in size (considering the
singles samples altogether) higher than most other studies of the \HV
network. The new aspects that make the present analysis unique are the
focus on the \MV and \LV network that no other study takes into
account, and also the usage of weighted graph that other studies
miss. In addition, the analysis is complete investigating betweenness
and resilience analysis through the computation of various failure
policies. Nonetheless the analysis examines the Small-world property
for the \MV and \LV network showing how the samples under exam do not
belong to this class of networks.

\section{Conclusions}\label{sec:conc}

When facing a global complex system such as the Power Grid, it is
necessary to combine precise tools to consider the local phenomena
with global statistical tools that provide an overall view. In the
present study, we look with a weighted model at the Medium and
Low Voltage Grids with the aim of understanding its potentials as
infrastructure for delocalized energy distribution. The
study takes the North Netherlands data as basis for its analysis
and proposes and adapts a number of statistical topological
measures for the specific goal. 

A number of novelties are a trademark of the current proposal:
the study of the lower layers of the Power Grid, the study of energy
distribution rather than simply resilience, the use of a weighted
topological model and, most notably, the proposal of tying the
topological properties to values denoting the attractiveness for the
prosumers to trade energy. But let us be more specific in summarizing
the results of the present treatment.

An interesting aspect is the difference experienced in the shortest
path between the weighted and unweighted graph. The difference is most
notable in the number of nodes that need to be
traversed. This is interesting since the
weighted paths considering resistance are closer to the ones really
travelled by the energy flow, therefore traversing much more nodes
than the ideal situation can lead to additional losses in substations
and transformers together with a larger number of potential point of
failure.

Considering graph statistical properties, the node degree distributions tend to
follow a power-law (at least for the most significant samples), that is there are few nodes that have many
connections, while the majority has a very limited number. As the
literature shows there is quite a common consensus about the type of
node degree distribution \HV Power Grids have, i.e., exponential,
while a study that takes into account many more
nodes obtains a power-law distribution~ \cite{Chassin05}. The explanation of the \MLV
Grids may reside in the relatively small number of stations and
transformers that receive their input from the \HV network and have to
distribute this input to many more substation at lower
voltages. 

The betweenness plots have a specific characteristic for the \LV samples:
unlike what presented in the literature reporting a power-law characterizing
the High Voltage, the \LV follows an exponential decay. The topology
itself of the \LV network, in which the paths are much more forced due
to the greater hierarchy of the \LV network, implies a betweenness
distribution with a more compressed tail than a fat-tailed
power-law. While a more meshed network as the \MV has more nodes that
take part in different number of paths.

Another remarkable aspect is the relatively higher tolerance
that is shown by the \MV network when edges are removed compared to the \LV network. This is due to the more meshed structure of the \MV which is
therefore less prone to failures than the Low Voltage. In fact, the average
number of links that need to be eliminated at the same time to cut the network in two
or more components are more than the ones for Low
Voltage.

From these findings, we move towards our claim, that is, the
influence of these global topological factors on the actual possible
use of the \PG as an infrastructure for delocalized energy exchange. We
make an initial proposal of how the \CNA topological metrics can be clustered in two
parameters $\alpha$ and $\beta$ that need to be optimized in order to
facilitate energy negotiation among prosumers. To optimize them,
one may have to change the topological properties of the
Grid. It might turn out that to reduce the WCPL needs to lay
more cables between rural and urbanized areas, or to put more
interconnections between neighboring urbanized areas.

Clearly there is much more to be investigated in the direction of
Complex Network Analysis for the Power Grid. The models need to be
enriched with ever more Power Grid specific characteristics. The
dynamics of the growth of the Power Grid may also provide new
insights. Simulations of possible future scenarios will also help
to identify the right infrastructure of the future (Smart) Grid, not to
mention the importance of coupling such technical studies with their
economic counterparts.

\section*{Acknowledgment}

The authors thank Hanzehogeschool Groningen in the person of Peter Kamphuis, Enexis B.V. in the person of Hemmo Hulzebos and Phase-to-Phase B.V. in the person of Berto Jansen for the Grid data and useful information. Pagani is supported by University of Groningen with the Ubbo Emmius
Fellowship 2009. The work is supported by the Dutch National Research
Council under the NWO Smart Energy Systems programme, contract
no.~600.647.000.004.

\bibliographystyle{IEEEtran}
\bibliography{andrea.bib}

\end{document}